\documentclass[pdflatex,sn-mathphys-num]{sn-jnl}

\usepackage{graphicx}%
\usepackage{multirow}%
\usepackage{amsmath,amssymb,amsfonts}%
\usepackage{amsthm}%
\usepackage{mathrsfs}%
\usepackage[title]{appendix}%
\usepackage{xcolor}%
\usepackage{textcomp}%
\usepackage{manyfoot}%
\usepackage{booktabs}%
\usepackage{algorithm}%
\usepackage{algorithmicx}%
\usepackage{algpseudocode}%
\usepackage{listings}%

\newcommand{\Lya}{Ly$\alpha$}
\newcommand{\OII}{[O\,{\sc ii}]}
\newcommand{\OIII}{[O\,{\sc iii}]}

\newcommand{\Ha}{H$\alpha$}
\newcommand{\Hb}{H$\beta$}

\newcommand{\CIII}{C{\sc iii}]}
\newcommand{\CIV}{C{\sc iv}}
\newcommand{\HeII}{He{\sc ii}}

\newcommand{\xiion}{$\xi_{\rm ion}$}

\newcommand{\ergscm}{erg\,s$^{-1}$\,cm$^{-2}$}

\newcommand{\ergHz}{erg$^{-1}$\,Hz}

\newcommand{\kms}{km\,s$^{-1}$}
\newcommand{\Oabundance}{$12+\log ({\rm O/H})$}

\newcommand{\HII}{H{\sc ii}}

\newcommand{\Msunyr}{$M_{\odot}\,{\rm yr}^{-1}$}
\newcommand{\Msun}{$M_{\odot}$}
\newcommand{\Mstar}{$M_{\star}$}

\newcommand{\MsunLsun}{$(M$/$L)_{\odot}$}

\newcommand{\Muv}{$M_{\rm UV}$}

\newcommand{\farcm}{.\kern -0.7ex\raisebox{.9ex}{\scriptsize$\prime$}}%
\newcommand{\farcs}{.\kern -0.7ex\raisebox{.9ex}{\scriptsize$\prime\prime$}}%
\newcommand{\farcstiny}{.\kern -0.7ex\raisebox{.9ex}{\tiny$\prime\prime$}}%

\theoremstyle{thmstyleone}%
\theoremstyle{thmstyletwo}%

\theoremstyle{thmstylethree}%

\usepackage{multibib}
\newcites{meth}{Methods References}
\newcites{supp}{Supplementary Information References}

\setcitestyle{super}
\setcitestyle{open={},close={}}

\begin{document}

\title[]{An Ultra-Faint, Chemically Primitive Galaxy Forming in the Reionization Era}


\author*[1,2,3]{\fnm{Kimihiko} \sur{Nakajima}}\email{knakajima@staff.kanazawa-u.ac.jp}

\author[3,4,5,6]{\fnm{Masami} \sur{Ouchi}}

\author[4]{\fnm{Yuichi} \sur{Harikane}}

\author[7]{\fnm{Eros} \sur{Vanzella}}

\author[4]{\fnm{Yoshiaki} \sur{Ono}}

\author[8,9,10]{\fnm{Yuki} \sur{Isobe}}

\author[5,3]{\fnm{Moka} \sur{Nishigaki}}

\author[3,5]{\fnm{Takuji} \sur{Tsujimoto}}

\author[3,5,11]{\fnm{Fumitaka} \sur{Nakamura}}

\author[4,11]{\fnm{Yi} \sur{Xu}}

\author[4,12]{\fnm{Hiroya} \sur{Umeda}}

\author[13]{\fnm{Yechi} \sur{Zhang}}

\affil*[1]{\orgdiv{Institute of Liberal Arts and Science}, \orgname{Kanazawa University}, \orgaddress{\street{Kakuma-machi}, \city{Kanazawa}, \postcode{920-1192}, \state{Ishikawa}, \country{Japan}}}

\affil[2]{\orgdiv{Division of Mathematical and Physical Sciences, Graduate School of Natural Science and Technology}, \orgname{Kanazawa University}, \orgaddress{\street{Kakuma-machi}, \city{Kanazawa}, \postcode{920-1192}, \state{Ishikawa}, \country{Japan}}}

\affil[3]{\orgname{National Astronomical Observatory of Japan}, \orgaddress{\street{2-21-1 Osawa}, \city{Mitaka}, \postcode{181-8588}, \state{Tokyo}, \country{Japan}}}

\affil[4]{\orgdiv{Institute for Cosmic Ray Research}, \orgname{The University of Tokyo}, \orgaddress{\street{5-1-5 Kashiwanoha}, \city{Kashiwa}, \postcode{277-8582}, \state{Chiba}, \country{Japan}}}

\affil[5]{\orgdiv{Department of Astronomical Science}, \orgname{SOKENDAI (The Graduate University for Advanced Studies)}, \orgaddress{\street{2-21-1 Osawa}, \city{Mitaka}, \postcode{181-8588}, \state{Tokyo}, \country{Japan}}}

\affil[6]{\orgdiv{Kavli Institute for the Physics and Mathematics of the Universe (WPI)}, \orgname{The University of Tokyo}, \orgaddress{\street{5-1-5 Kashiwanoha}, \city{Kashiwa}, \postcode{277-8582}, \state{Chiba}, \country{Japan}}}

\affil[7]{\orgdiv{Astrophysics and Space Science Observatory Bologna (OAS)}, \orgname{Istituto Nazionale di Astrofisica (INAF)}, \orgaddress{\street{Via Gobetti 93/3}, \postcode{40129}, \state{Bologna}, \country{Italy}}}

\affil[8]{\orgdiv{Kavli Institute for Cosmology}, \orgname{University of Cambridge}, \orgaddress{\street{Madingley Road}, \city{Cambridge}, \postcode{CB3 0HA}, \country{UK}}}

\affil[9]{\orgdiv{Cavendish Laboratory}, \orgname{University of Cambridge}, \orgaddress{\street{19 JJ Thomson Avenue}, \city{Cambridge}, \postcode{CB3 0HE}, \country{UK}}}

\affil[10]{\orgdiv{Waseda Research Institute for Science and Engineering, Faculty of Science and Engineering}, \orgname{Waseda University}, \orgaddress{\street{3-4-1 Okubo}, \city{Shinjuku}, \postcode{169-8555}, \state{Tokyo}, \country{Japan}}}

\affil[11]{\orgdiv{Department of Astronomy, Graduate School of Science}, \orgname{The University of Tokyo}, \orgaddress{\street{7-3-1 Hongo}, \city{Bunkyo}, \postcode{113-0033}, \state{Tokyo}, \country{Japan}}}

\affil[12]{\orgdiv{Department of Physics, Graduate School of Science}, \orgname{The University of Tokyo}, \orgaddress{\street{7-3-1 Hongo}, \city{Bunkyo}, \postcode{113-0033}, \state{Tokyo}, \country{Japan}}}

\affil[13]{\orgdiv{IPAC}, \orgname{California Institute of Technology}, \orgaddress{\street{MC 314-6, 1200 E. California Boulevard}, \city{Pasadena}, \postcode{91125}, \state{CA}, \country{USA}}}


\abstract{
The formation of the first stars and galaxies marked the onset of chemical enrichment, yet direct observations of such primordial systems remain elusive. Here we present James Webb Space Telescope spectroscopic observations of LAP1-B, an ultra-faint galaxy at redshift $z_{\rm spec}=6.625 \pm 0.001$, corresponding to a cosmic age of 800 million years after the Big Bang, strongly magnified by gravitational lensing. LAP1-B exhibits a gas-phase oxygen abundance of $(4.2\pm 1.8) \times 10^{-3}$ times the solar value, making it the most chemically primitive star-forming galaxy discovered to date. The galaxy displays an exceptionally hard ionizing radiation field, which is inconsistent with chemically enriched stellar populations or accreting black holes but matches theoretical predictions for an exceptionally metal-deficient stellar population \cite{NM2022}. It also shows an elevated carbon-to-oxygen abundance ratio for its metallicity in the interstellar medium, consistent with nucleosynthetic yields from a stellar population formed in the absence of initial metals \cite{HW2010,vanni2023,dEugenio2024}. The lack of detectable stellar continuum constrains the stellar mass to $<3,300$\,\Msun, while the dynamical mass, derived from emission-line kinematics, exceeds the combined stellar and gas mass and indicates a dominant dark matter halo. 
Our findings establish LAP1-B as a ``fossil in the making'', a direct high-redshift progenitor of the ancient ultra-faint dwarf galaxies observed in the local Universe, offering a rare window into the earliest stages of galaxy formation.
}

\maketitle

Understanding how the first stars and galaxies formed remains one of astronomy's most fundamental challenges. These early cosmic structures seeded the chemical and structural evolution of the Universe, producing the first heavy elements and giving rise to the diverse galaxy populations observed today, from massive systems like the Milky Way to faint, ancient relics such as ultra-faint dwarf galaxies (UFDs). The new generation of powerful telescopes, most notably the James Webb Space Telescope (JWST), has opened an unprecedented window into this early epoch, enabling the exploration of galaxies formed within a few hundred million years after the Big Bang \cite{curtis-lake2023_z13, harikane2024_spechighz, carniani2024_z14, naidu2026_z14p44}.

\begin{figure}[t!]
    \begin{center}
     \includegraphics[bb=0 0 709 319, width=1.\columnwidth]{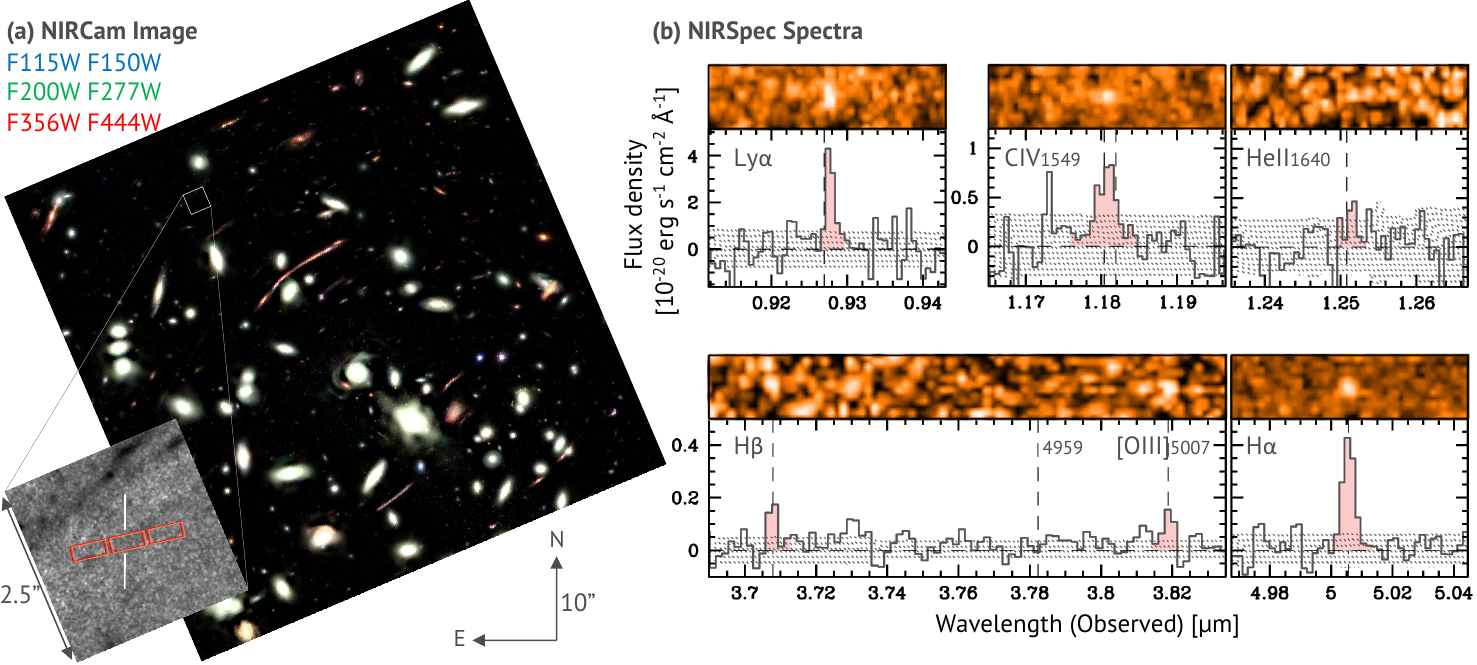}
    \caption{%
    	\textbf{NIRCam image and NIRSpec spectra of LAP1-B.}
	(\textbf{a Left:}) False-color NIRCam image of the MACS J0416 cluster, combining F115W and F150W (blue), F200W and F277W (green), and F356W and F444W (red). The inset shows a 2\farcstiny5$\times$2\farcstiny5 zoom-in of the full-composite image (stacked across F115W to F444W) centered on the location of LAP1-B (indicated with white lines), where no counterpart is detected. The two footprints of the NIRSpec 3-shutter slitlets used for spectroscopy are overlaid in red and orange.
	(\textbf{b Right:}) JWST/NIRSpec spectra of LAP1-B centered on the emission lines discussed in this work. In each panel, the top shows the 2D spectrum, and the bottom shows the extracted 1D spectrum (black) with the $1\sigma$ uncertainty shaded. The expected wavelengths of emission lines at the systemic redshift $z = 6.625$ are marked by vertical dashed lines. Emission signals detected within $\pm 1.5\times$ the instrumental FWHM are highlighted in red.
    }
    \label{fig:spec_snapshot}
    \end{center} 
\end{figure}

A primary objective in this new era is to observe the transition from pristine environments to the first stages of chemical enrichment. This search is largely motivated by the quest for the Universe's first-generation, metal-free stars known as Population III stars. However, identifying the galaxies that hosted these stars remains difficult due to their faintness. Indeed, early spectroscopic surveys with JWST have revealed a bias toward evolved systems, with none at $z=4$--$10$ confirmed below a ``metallicity floor'' of \Oabundance\ $\simeq 7.0$ ($\sim 2$\%\ solar) \cite{nakajima2023_jwst,curti2024_MZR}. Given these challenges, a promising avenue moving forward with JWST lies in targeting intrinsically fainter, emission-line dominated galaxies, especially those magnified by gravitational lensing, offering access to the most chemically primitive systems that would otherwise remain undetectable.

After decades of theoretical predictions and observational efforts, recent JWST investigations have begun probing this primitive frontier. One study \cite{maiolino2024_popIII} identified \HeII\ emission near the luminous $z=10.6$ galaxy GN-z11 \cite{oesch2016, bunker2023_gnz11}, a key signature indicative of the hard radiation field expected from Population III stars \cite{schaerer2003,inoue2011_metal_poor,NM2022}. However, the absence of additional lines, including metal lines, has so far prevented a firm determination of the source's nature and metallicity. 
A complementary photometric approach has identified extremely metal-poor candidates at $z=4$--$7$ based on strong hydrogen and weak metal line signatures \cite{nishigaki2023, fujimoto2025, hsiao2025, trussler2023}, with spectroscopic follow-up now underway \cite{hsiao2025}. However, bridging the gap between detecting metal-poor candidates and confirming their primitive nature remains a challenge. Critically, no system has yet demonstrated the dual hallmarks of primitive populations: extreme metal deficiency and an exceptionally hard ionizing spectrum.

Building on these developments, a compelling candidate for a chemically primitive system, LAP1-B, was recently identified at $z=6.6$ behind the gravitational lensing cluster MACS J0416 \cite{vanzella2023_metalpoor}. This object, magnified by a factor of $\mu \simeq 100$, is notable for its striking lack of detectable metal lines despite clear hydrogen features. In this study, we propose that the extreme metal deficiency and low stellar mass of LAP1-B signify a system caught at the threshold of chemical enrichment. By characterizing its ionizing spectrum, elemental abundances, and kinematic properties, we evaluate LAP1-B as a ``fossil in the making'', a direct high-redshift progenitor of the ancient UFDs observed in the local Universe.

Here we present results from very deep, medium-resolution JWST spectroscopy using Near Infrared Spectrograph (NIRSpec) \cite{jakobsen2022} of LAP1-B, an ultra-faint galaxy with a rest-frame ultraviolet absolute magnitude fainter than \Muv\ $=-10.4$ ($3\sigma$). The observations, conducted on 4-5 November 2024, covered wavelength ranges $0.8$--$1.8\,\mu$m and $2.9$--$5.2\,\mu$m with $16.37$ hours total integration time per setting. Data reduction followed procedures described in detail in the Methods section. Fig.~\ref{fig:spec_snapshot} presents the one-dimensional (1D) and two-dimensional (2D) JWST/NIRSpec spectra of LAP1-B. At redshift $z=6.625 \pm 0.001$, the medium-resolution data reveal five significant emission lines: \Ha, \Hb, \OIII$\lambda 5007$, \CIV$\lambda 1549$, and \Lya.

A key result is the weak but significant detection of metal lines, specifically \OIII\ and \CIV, at the position of LAP1-B. These features suggest the interstellar medium (ISM) has been enriched by only a minimal number of stellar generations in the galaxy. 
In particular, the \OIII$\lambda 5007$$/$\Hb\ flux ratio serves as a widely used proxy for oxygen abundance (O$/$H) in ionized nebulae in the ISM \cite{MM2019}. Our measurement of \OIII$/$\Hb~$= 0.69 \pm 0.28$ falls in an extremely low-metallicity regime, forcing us to extrapolate theoretical models beyond their empirical calibration limit of \Oabundance\ $\simeq 7.0$. Assuming a high ionization parameter consistent with such metal-poor environments (see Methods and Extended Data Fig.~\ref{fig:Z_R3}), this procedure yields an oxygen abundance of \Oabundance\ $=6.31^{+0.15}_{-0.23}$, representing $(4.2\pm 1.8)\times 10^{-3}$ of solar metallicity. 
This confirms LAP1-B as among the most chemically primitive star-forming galaxies identified at any epoch to date. Consistent with its primitive nature, the galaxy shows negligible dust extinction, as evidenced by an \Ha/\Hb\ ratio that matches the theoretical value for Case B recombination.

\begin{figure}[t!]
    \begin{center}
    \includegraphics[bb=0 0 549 363, width=1.\columnwidth]{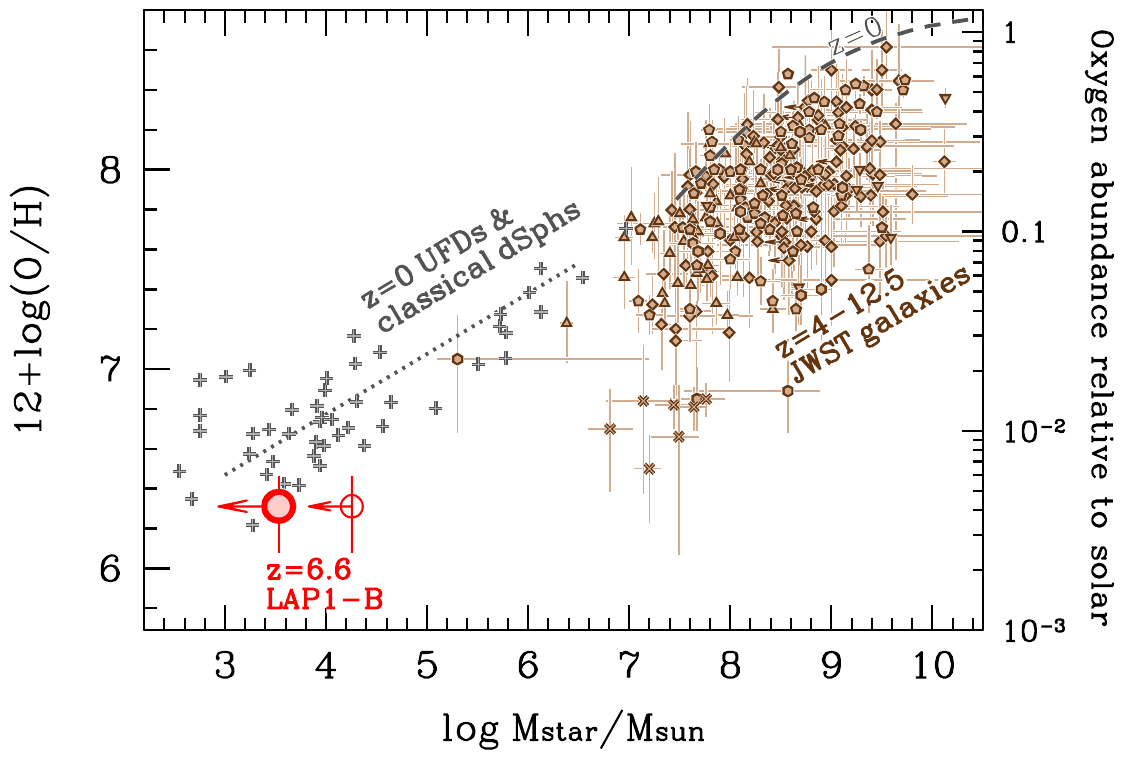}
    \caption{%
	\textbf{Chemical enrichment as a function of stellar mass across cosmic time.}
	The oxygen abundance and stellar mass of LAP1-B are shown with two symbols. The filled red circle represents our fiducial measurement, while the open red circle indicates the conservative upper limit on mass, as discussed in the text.
	This measurement is compared with those of $300$ other galaxies at $z=4$--$12.5$ identified with JWST 
	(brown symbols), whose metallicities were derived in a comparable manner. 
	For local comparison, we show $z=0$ UFDs and more massive counterparts of UFDs, referred to as classical dwarf spheroidal galaxies (dSphs) as grey crosses (individual systems) and a grey dotted line (average relation). Their metallicities are based on stellar [Fe/H] measurements derived from the metallicity distributions of individual stars. These values are converted to oxygen abundances assuming [O/Fe] $=+0.5$.
	We note this is an approximate comparison; the present-day stellar masses of these dwarf galaxies can be lower limits on their progenitor masses due to stellar evolutionary mass loss, and the precise behavior of the [O/Fe] ratio in such metal-poor environments is a subject of ongoing research.
	References for the data compiled here are provided in the Methods section.
    }
    \label{fig:MZR}
    \end{center} 
\end{figure}

Combined with a stellar mass constraint of \Mstar\ $<3,300$\,\Msun\ ($3\sigma$) from non-detection in deep JWST/NIRCam imaging (see Methods), LAP1-B occupies an extreme position on the stellar mass-metallicity relation (Fig.~\ref{fig:MZR}). 
This conclusion remains robust even when adopting a more conservative mass limit of \Mstar\ $<1.8\times 10^4$\,\Msun\ ($3\sigma$; see open symbol in Fig.~\ref{fig:MZR}).
Thanks to the advent of JWST, recent studies have begun to map the relation in the distant Universe, up to redshift $z=10$--$12.5$ \cite{alvarez-marquez2025, hsiao2024, zavala2025}, with oxygen abundance commonly used as a proxy for total metallicity. In Fig.~\ref{fig:MZR}, we compile a sample of $300$ high-redshift galaxies at $z=4$--$12.5$ (brown symbols), most of which have stellar masses exceeding $10^7$\,\Msun. By comparison, LAP1-B is over $1,000$ times less massive yet falls within the expected metallicity range when the relation is extrapolated to these low-mass extremes. 
LAP1-B thus represents a unique laboratory for studying star formation in a regime of extreme low mass and metal deficiency that has remained entirely inaccessible.

\begin{figure}[t!]
  \centering
    \begin{tabular}{c}
      \begin{minipage}{0.485\hsize}
        \begin{center}
         \includegraphics[bb=0 0 532 470, width=1.\textwidth]{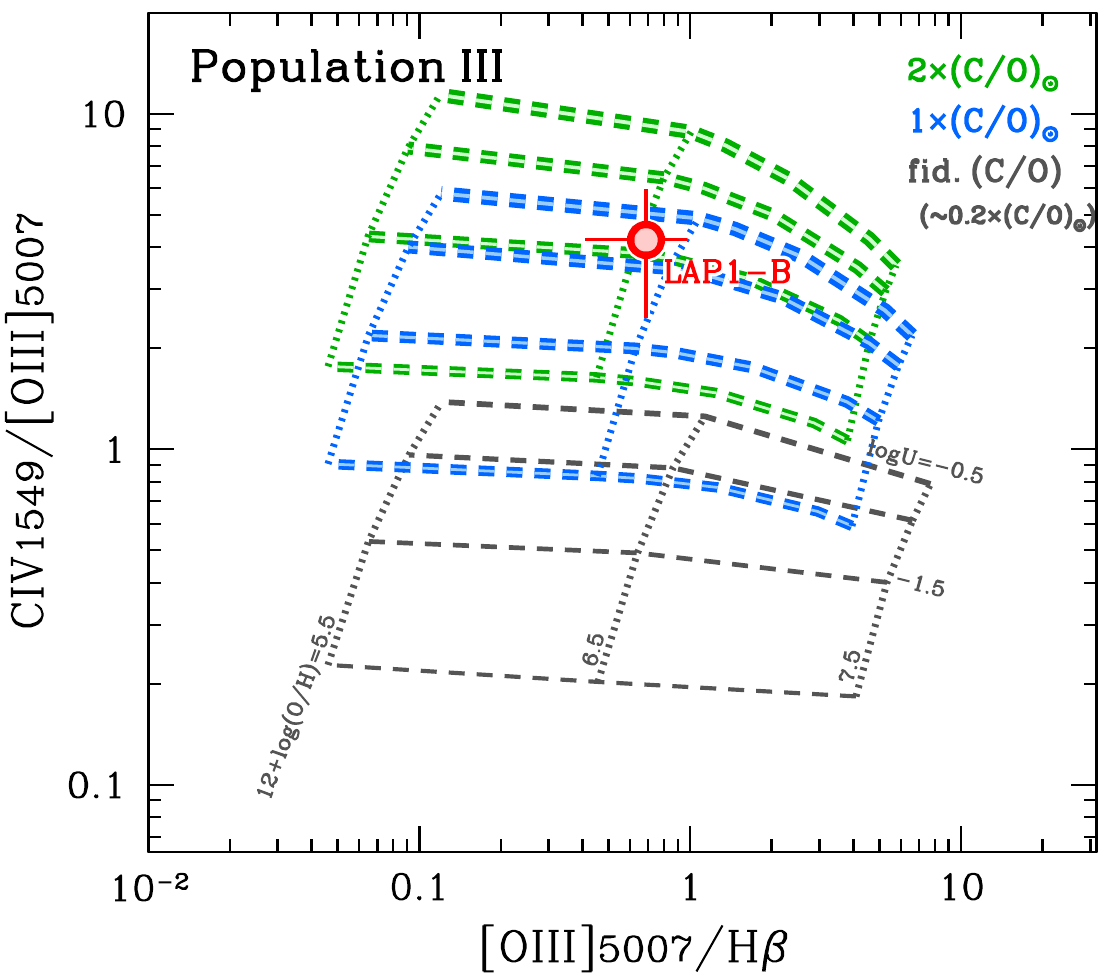}
        \end{center}
      \end{minipage}
      \begin{minipage}{0.485\hsize}
        \begin{center}
         \includegraphics[bb=0 0 532 470, width=1.\textwidth]{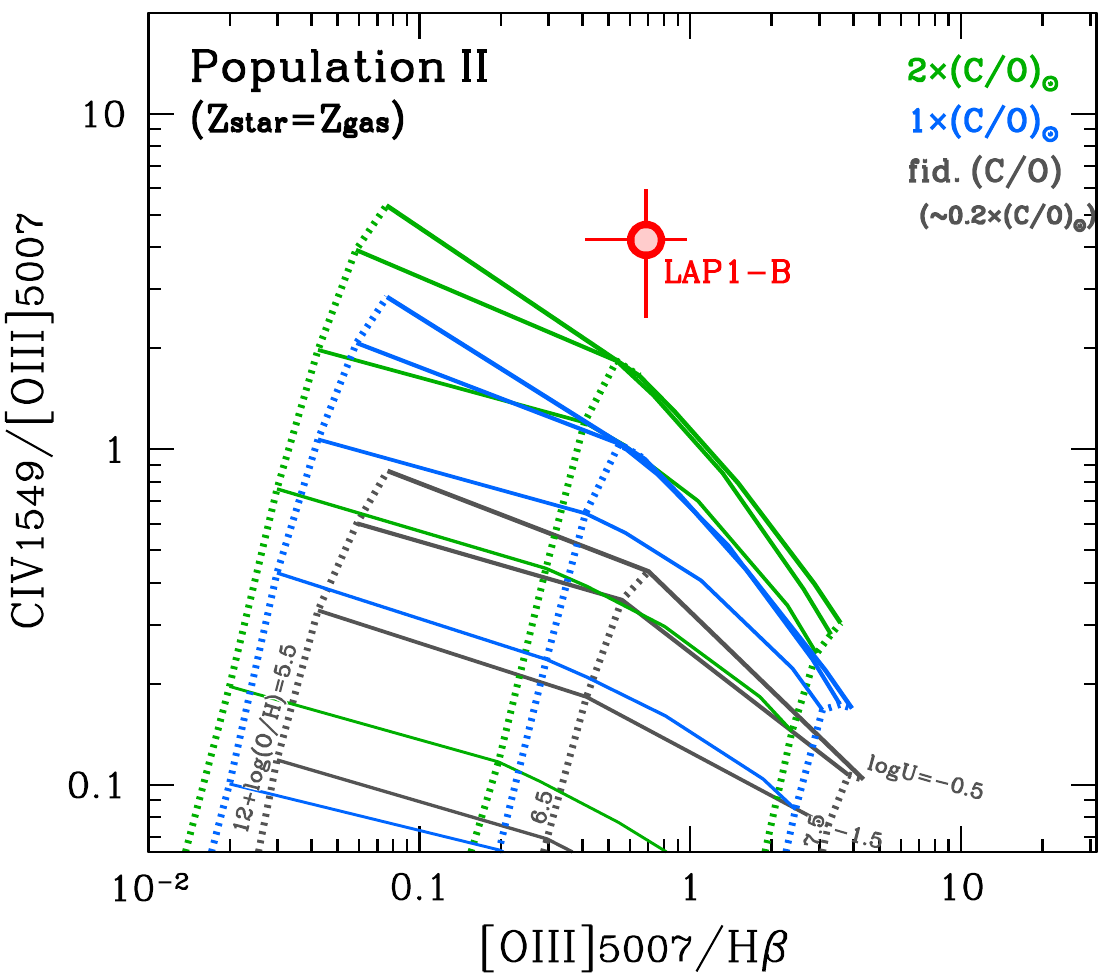}
        \end{center}
      \end{minipage}
      \\
      \\
      \begin{minipage}[b]{0.32\hsize}
        \begin{center}
          \includegraphics[bb=0 0 532 470, width=1.\columnwidth]{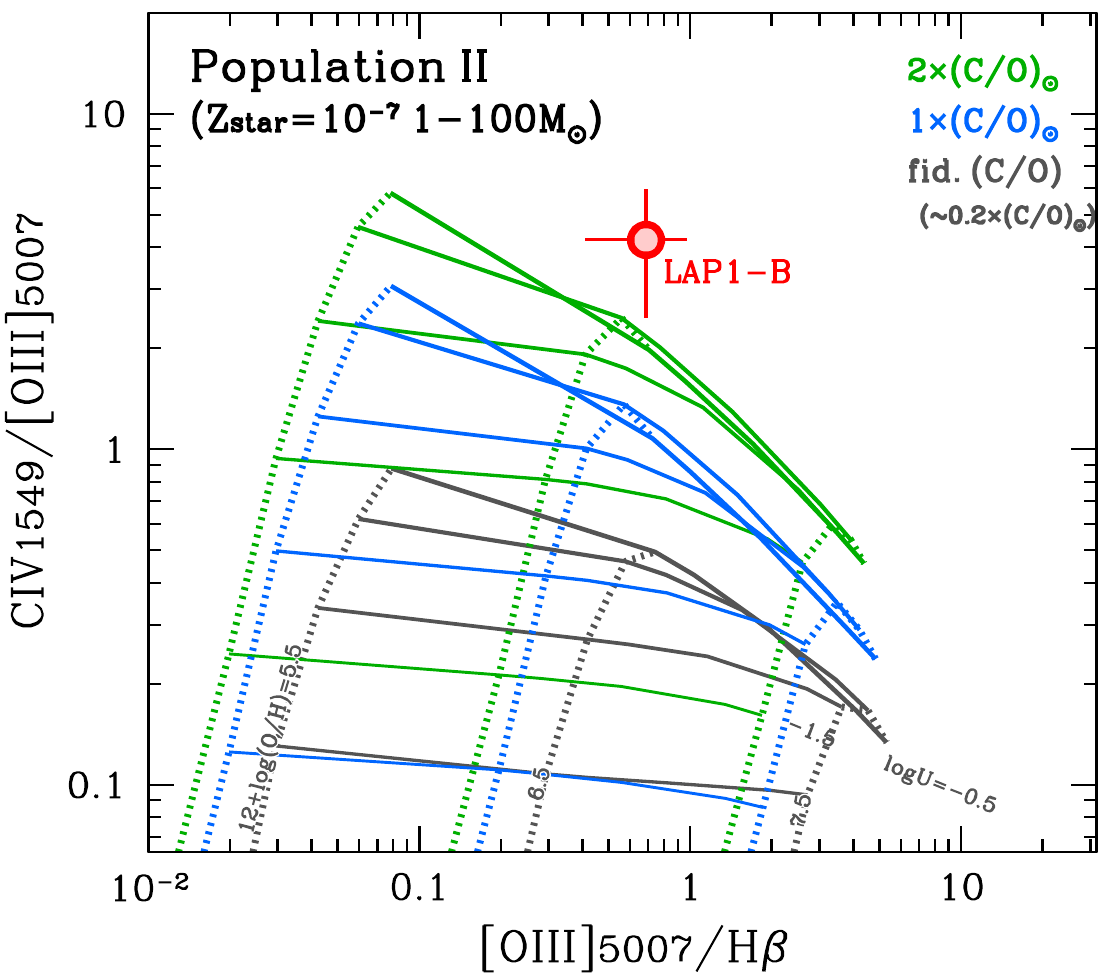}
        \end{center}
      \end{minipage}
      \begin{minipage}[b]{0.32\hsize}
        \begin{center}
         \includegraphics[bb=0 0 532 470, width=1.\columnwidth]{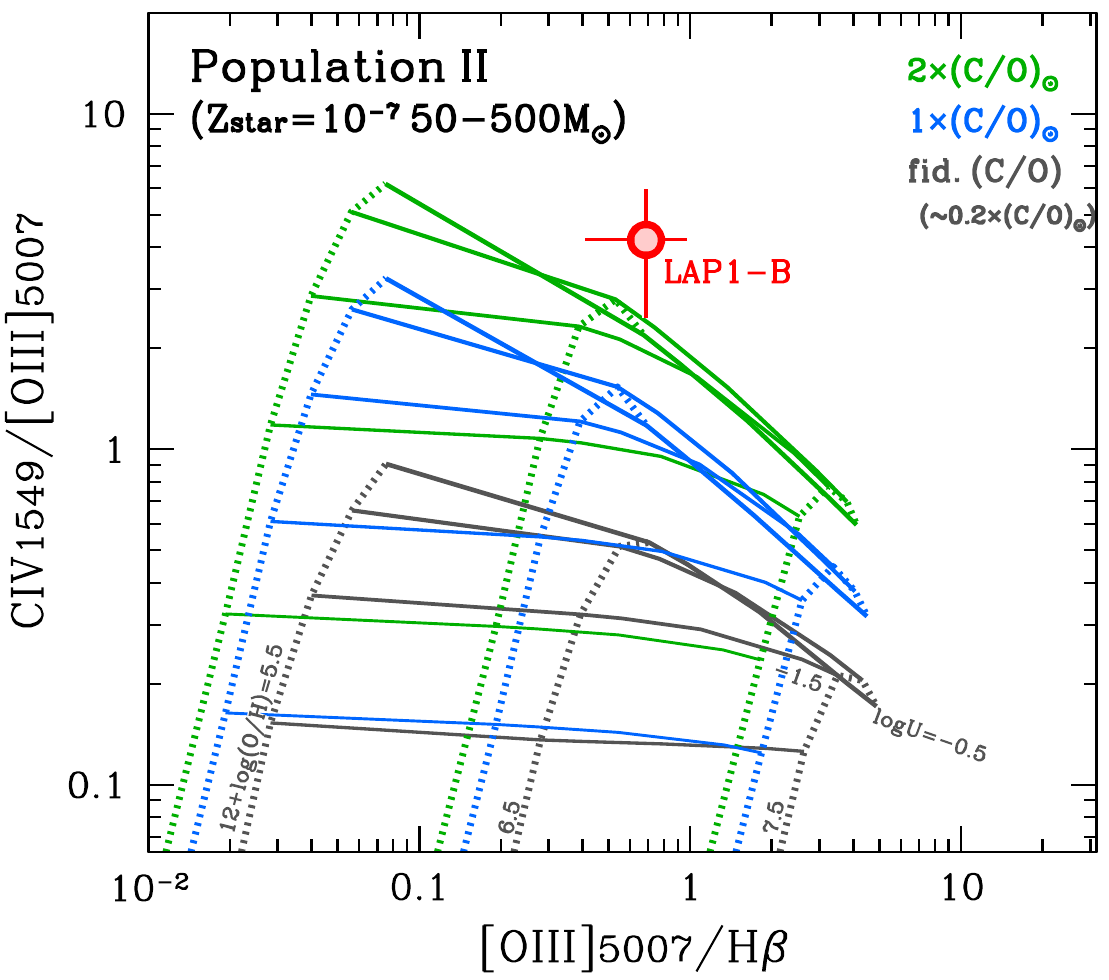}
        \end{center}
      \end{minipage}
      \begin{minipage}[b]{0.32\hsize}
        \begin{center}
         \includegraphics[bb=0 0 532 470, width=1.\columnwidth]{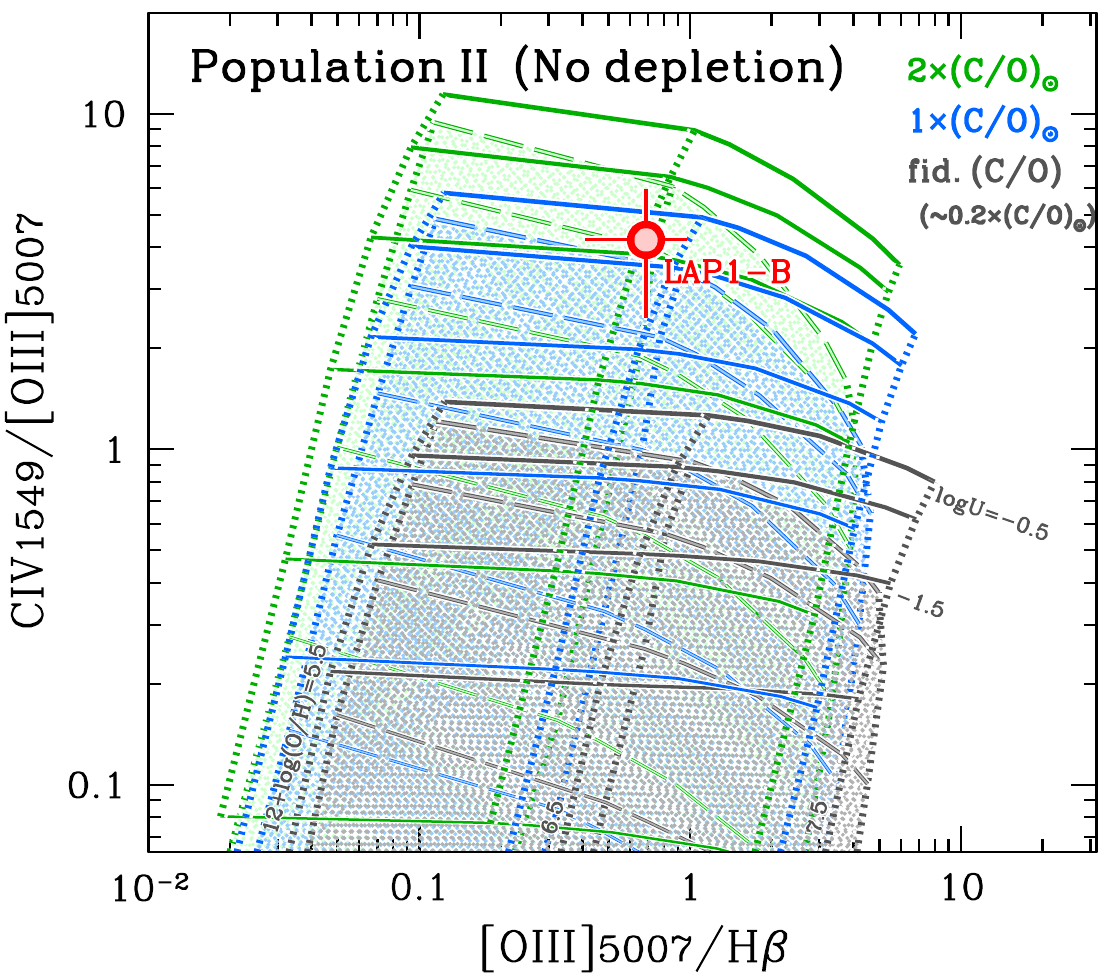}
        \end{center}
      \end{minipage}
    \end{tabular}
    \caption{%
    	\textbf{Comparison of observed emission-line ratios with photoionization models.}
	The observed line ratios for LAP1-B (red circle) are compared to photoionization model predictions \cite{NM2022} in the \CIV/\OIII\ versus \OIII/\Hb\ diagnostic diagram. The top row contrasts fiducial models: a zero-metallicity Population III population (left) versus a chemically enriched Population II population with standard assumptions (matched stellar and gas-phase metallicities, Kroupa IMF; right). The bottom row explores variations of the Population II models to test the impact of IMF and dust depletion. The first two bottom panels assume an extremely low stellar metallicity ($Z_{\text{star}}=10^{-7}$) and compare a standard IMF ($1$--$100$\,\Msun; left) to one composed of very massive stars ($50$--$500$\,\Msun; center). The bottom-right panel tests the role of dust depletion by assuming zero depletion for two scenarios: the fiducial Population II model ($Z_{\text{star}} = Z_{\text{gas}}$; long-dashed with shade) and the extreme top-heavy model ($Z_{\text{star}} = 10^{-7}, 50$--$500$\,\Msun; solid lines).
	Each panel illustrates model grids for three C/O abundance ratios: $2\times$ solar (green), solar (blue), and an empirically motivated relation (grey) corresponding to $\sim 0.2\times$ solar at these metallicities. Grids connect points of constant oxygen abundance and ionization parameter, with values indicated in the legend.
	}
    \label{fig:c4o3_o3hb}
\end{figure}

Beyond its extreme metallicity, LAP1-B notably displays an exceptionally hard ionizing spectrum, with an ionizing photon production efficiency of $\log$\,\xiion ($/$\ergHz) $> 26.1$ ($3\sigma$). This extreme value is incompatible with standard metal-enriched stellar populations or black hole accretion, aligning only with metal-free (Population III) stars or extreme Population II models ($Z=10^{-7}$) characterized by an initial mass function (IMF) composed exclusively of very massive stars (Extended Data Fig.~\ref{fig:diagnostics_popIII} right; see also refs \cite{vanzella2024_T2c, schaerer2025}). Consistent with these primitive scenarios, the system also places a strong lower limit on the equivalent width, EW(\Ha) $>1800$\,\AA\ ($3\sigma$), which further disfavors a dominant black hole contribution. The non-detection of the key diagnostic line \HeII$\lambda$1640 is also consistent with these specific stellar models (Extended Data Fig.~\ref{fig:diagnostics_popIII} left).

The detection of \CIV\ emission reinforces the presence of this intense radiation field,
as its production requires high-energy photons ($>47.9$\,eV). As shown in Fig.~\ref{fig:c4o3_o3hb}, the observed \CIV/\OIII\ and \OIII/\Hb\ line ratios are well reproduced by Population III models with a carbon-to-oxygen (C/O) abundance ratio of $1$--$2$ times the solar value (top left panel). 
The comparison with Population II models is more complex, as this diagnostic is highly sensitive to the assumed elemental depletion factors, which remain uncertain in a (near) dust-free environment like LAP1-B. While standard Population II models, even those with extremely low metallicity, fail to reproduce the data (top right and bottom two left panels) as long as typical depletions are assumed, a model assuming zero depletions can reproduce the observed \CIV/\OIII\ ratio (bottom right). 
Consequently, while the \CIV\ detection alone cannot distinguish these scenarios, its presence combined with the large \xiion\ firmly establishes LAP1-B as a primitive star-forming system dominated by hard ionizing radiation.

\begin{figure}[t!]
    \begin{center}
    \includegraphics[bb=0 0 549 363, width=1.\columnwidth]{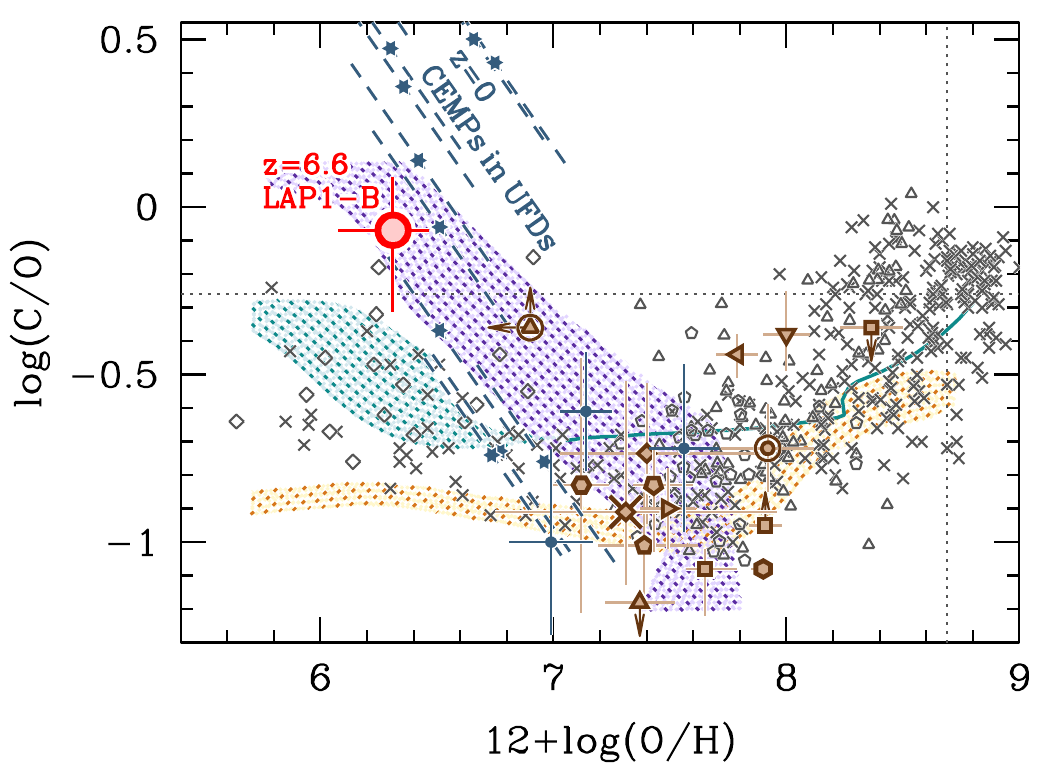}
    \caption{%
    	\textbf{Carbon-to-oxygen abundance ratios.}
	The C/O ratio of LAP1-B (red) is plotted as a function of oxygen abundance,
	falling above the solar C/O value (horizontal dotted line) despite its extremely low metallicity.
	This measurement is compared with a compilation of other astrophysical objects: high-redshift JWST galaxies (brown symbols), Galactic halo and disk stars (grey crosses), damped Lyman-$\alpha$ systems (grey diamonds), and nearby/intermediate-redshift galaxies (grey triangles/pentagons). To establish a local fossil-record comparison, we plot individual stars from present-day UFDs: blue-grey filled circles denote stars with direct C and O measurements, while blue-grey hexagrams represent CEMP stars. For the CEMP population, C/O and O/H ratios are inferred using the empirical relationship between [O/Fe] and [Fe/H] observed in Galactic halo CEMP-no stars, a population thought to originate from massive Population III stars (see Methods). Diagonal error bars (dashed blue-grey) account for the coupled uncertainty inherent in this conversion.
	Theoretical predictions are overlaid for comparison. Purple shaded regions represent enrichment from Population III stars and faint supernovae (adapted from ref \cite{dEugenio2024} based on ref \cite{vanni2023}). Blue-green tracks show chemical evolution models \cite{CL2002}, where the hatched area brackets predictions for different Population III IMFs. Orange shaded regions represent fiducial Population II enrichment models \cite{kobayashi2020}. While the evolution tracks (blue-green and orange) are calculated for Milky Way-like systems, they serve as a baseline for distinguishing primordial and standard chemical enrichment pathways. 
	References for all compiled data are provided in the Methods.
	}
    \label{fig:co_oh}
    \end{center} 
\end{figure}

The carbon-to-oxygen abundance ratio in LAP1-B offers complementary insight into its past star-formation and chemical enrichment history, as the ISM composition encodes the nucleosynthetic yields of earlier stellar populations. Assuming an electron temperature consistent with the system's extremely low metallicity, and that carbon resides predominantly in the triply ionized state (see Methods), we derive $\log$(C/O) $= -0.07^{+0.16}_{-0.24}$ from the observed line ratio \CIV$/$\OIII\ $= 4.2 \pm 1.8$. This corresponds to a C/O ratio of $1$--$2$ times the solar value, in agreement with photoionization models (Fig.~\ref{fig:c4o3_o3hb}).
As shown in Fig.~\ref{fig:co_oh}, LAP1-B lies above the C/O trend observed in Galactic metal-poor stars and damped Lyman-$\alpha$ systems at similar oxygen abundances (\Oabundance\ $\lesssim 7$), where C/O typically increases with decreasing O/H. This elevated C/O ratio distinguishes LAP1-B from systems shaped by conventional enrichment pathways.

Comparison with chemical evolution models highlights the specific nature of this anomaly. Standard Population II models (shown in orange), while accurate at higher metallicities, fail to reproduce the data, predicting a C/O plateau rather than the observed elevation \cite{MM2002, akerman2004, kobayashi2020}.
In contrast, enrichment by metal-free Population III stars offers a compelling explanation (shown in purple and blue-green). 
The absence of initial metals leads to more compact stellar structures and strongly suppressed mass loss, fundamentally altering stellar evolution and the final carbon–oxygen core structure \cite{schaerer2002,HW2002,woosley2002}. In such primordial environments, these compact Population III stars are thought to be more susceptible to ``faint'' supernova explosions characterized by substantial fallback \cite{UN2002,UN2003,iwamoto2005,HW2010,ishigaki2014}. In these events, the low explosion energy is insufficient to eject the oxygen-rich inner layers, which instead collapse into a central remnant, while the carbon-rich outer layers are preferentially expelled into the surrounding ISM.

While the precise progenitor properties remain debated, the abundance pattern strongly points to an enrichment history dominated by the legacy of recent Population III supernovae (purple region \cite{vanni2023}). Attributing this to a single burst of massive stars is theoretically possible but requires fine-tuned conditions, including synchronized explosions, efficient fallback, and non-standard energy injection \cite{HW2010, ji2024}. Alternatively, the observed chemistry is equally consistent with stochastic enrichment by previous generations of faint Population III stars. In either framework, LAP1-B serves as a fossil record of the discrete enrichment events that defined the earliest stages of galaxy assembly.

The combined evidence: an exceptionally hard ionizing spectrum, extremely low stellar mass, and a minimally enriched ISM, points to a galaxy observed in a brief but critical stage of its formation.
In this framework, the gas was likely pre-enriched by a previous generation of Population III stars, which left behind an elevated C/O ratio, while the ionizing flux is currently provided by a subsequent, or perhaps coeval, generation of an extremely metal-deficient stellar population (cf. refs \cite{katz2023_popIII, rusta2025}).
This definitively stellar origin contrasts sharply with a metal-poor system at $z=7$ that is powered by an accreting massive black hole \cite{maiolino2025_metalpoorBH}, providing a unique and direct view of the immediate aftermath of the first enrichment events in the Universe. 
Consistent with simulations of delayed star formation (see Methods),  LAP1-B represents a critical ``missing link'' in the transition from the cosmic dawn to the established galaxy populations of the early Universe.

Beyond its stellar and chemical properties, the internal kinematics of LAP1-B provide critical insight into its mass composition. The \Ha\ emission line exhibits a modest but significant broadening beyond the instrumental resolution, yielding a velocity dispersion of $58.3 \pm 17.8$\,\kms.
Using the virial theorem and assuming an intrinsic size of $\simeq 10$ parsec, estimated from de-lensed Balmer emission map \cite{vanzella2023_metalpoor}, we infer a dynamical mass of order $10^7$\,\Msun\ (see Methods). This inferred dynamical mass exceeds the combined stellar and gas mass estimates, \Mstar\ $<3,300$\,\Msun\  ($3\sigma$) and $M_{\rm gas} \sim$ a few $\times 10^5$\,\Msun\ (see Methods), by more than two orders of magnitude. The resulting baryon fraction is $\lesssim 1$\%, reinforcing the interpretation that LAP1-B is embedded within a dominant dark matter halo.

Such a dark matter-dominated mass budget is a hallmark feature of UFDs in the present-day Universe \cite{simon2019}, suggesting a potential evolutionary link between LAP1-B and these ancient relic systems. 
This link is solidified by a remarkable chemical similarity. Despite the known complexities in comparing gas-phase and stellar abundances (see Methods), we find that LAP1-B lies directly on the well-established mass-metallicity relation defined by UFDs (Fig.~\ref{fig:MZR} at \Mstar\ $<10^5$\,\Msun). Furthermore, LAP1-B shares the characteristic carbon-enhanced signature found in individual stars within these systems. As shown in Fig.~\ref{fig:co_oh}, LAP1-B's C/O and O/H ratios directly align with those of Carbon-Enhanced Metal-Poor (CEMP) stars identified in local UFDs. The distinct chemical signatures in UFDs are themselves widely interpreted as the imprint of enrichment from Population III supernovae \cite{jeon2017, rossi2025, heiger2025}. This aligns perfectly with our independent interpretation of LAP1-B, suggesting that its ISM represents a direct snapshot of the gas from which the stars in today's UFDs were born. 
Taken together, these similarities support the view that LAP1-B represents a formative phase of a UFD in the early Universe, captured while star formation was still ongoing.

This scenario is consistent with the prevailing view that today's UFDs largely formed at the epoch of reionization, which concluded around $z \simeq 5.3$ \cite{bosman2022}. Reionization likely quenched star formation in low-mass halos by heating the intergalactic medium and suppressing gas accretion onto shallow gravitational potentials \cite{bovill2009, salvadori2009}. 
LAP1-B therefore provides a rare opportunity to bridge the observational gap between early star-forming systems and local relic dwarfs, capturing a galaxy in active formation before reionization quenched its star-forming activity.


\bibliography{Refs_sn-article}{}


\newpage

\section*{Methods}
\label{sec:methods}

\subsection*{Adopted cosmology and conventions}
\label{ssec:cosmology_conventions}

Throughout the paper we adopt a standard $\Lambda$CDM cosmology with $\Omega_{\Lambda}=0.7$, $\Omega_{m}=0.3$, and $H_0=70$\,\kms\,Mpc$^{-1}$. All magnitudes are given in the AB system \citemeth{OG1983}
As for the chemical abundance notation, the square brackets indicate dex units, meaning decimal exponent defined here with respect to the Sun. The solar chemical compositions reported in ref \citemeth{asplund2009} are adopted throughout this paper.

\subsection*{JWST/NIRSpec observations and data reduction}
\label{ssec:observations}

The LAP1 system, located in the lensing cluster field MACS J0416, comprises a strongly magnified \Lya-emitting arc at redshift $z=6.6$ \citemeth{vanzella2020}. Initial JWST observations \cite{vanzella2023_metalpoor} revealed three distinct clumps in the Balmer emission line maps. These are interpreted, using the latest gravitational lensing model \cite{vanzella2023_metalpoor}$^{,}$\citemeth{bergamini2023}, as arising from two circular components, A and B. Component A forms a pair of unresolved images near the critical curve (A1 and A2), while Component B produces mirrored features (B1 and B2). The \OIII\ line maps indicate that most of the metal-line emission arises from Component A, suggesting that Component B is markedly more metal-poor. In this study, we focus on Component B1, referred to throughout as LAP1-B.

Deriving the intrinsic properties of this complex system requires a detailed lens model. We adopt the model developed specifically for LAP1 \cite{vanzella2023_metalpoor}. Their analysis refines the state-of-the-art cluster mass model using the high-precision JWST astrometry of the lensed knots \citemeth{bergamini2023}. Through a sophisticated forward-modeling approach, this work derives a magnification factor of $\mu=98^{+5}_{-4}$, which we use to correct all derived quantities in this paper.
We emphasize that our primary conclusions regarding the gas-phase chemistry and ionization state, which point to a primordial origin, are unaffected by lensing uncertainties, as they are derived from emission line ratios that are independent of the magnification.

Observations of LAP1-B were conducted with the JWST/NIRSpec Micro-Shutter Assembly (MSA) on November 4-5, 2024, under program GO-4750 (PI: K. Nakajima). This program, titled "Deep Reconnaissance of Early Assemblies of Metal-poor Star formation" (DREAMS), will be detailed in a forthcoming publication (Nakajima et al., in preparation).
We employed the medium-resolution grating configurations G140M/F070LP and G395M/F290LP, which provide spectral resolution $R \sim 1,000$. Observations were executed across eight observing blocks, four with MSA configuration 3001 and four with 4001, using a 3-shutter slitlet pattern. Each block included three exposures per grating, with individual exposure times of 2947\,sec. Due to slitlet placement constraints imposed by nearby clumps of LAP1, LAP1-B was only covered by two of the three exposures in half of the nodding blocks. The total integration time for LAP1-B was 16.37 hours per grating. The IRS$^2$ readout mode was used. The locations of the two NIRSpec slitlets for LAP1-B are shown in Fig.~\ref{fig:spec_snapshot}(a), with the 3001 configuration in red and the 4001 configuration in orange.

Data were reduced using the JWST calibration pipeline (version 1.17.1) under CRDS context jwst\_1298.pmap provided by STScI. Standard steps included detector-level calibration, wavelength and flux calibration, and background subtraction. Crucially, for background subtraction, we used only one of the two alternative nodded slitlet positions rather than combining them. This was necessary to avoid self-subtraction or contamination from nearby Component A, which lies at the same redshift. In each case, we explicitly selected the nod position free from potential contamination.
Although the F070LP filter is nominally truncated at $1.27\,\mu$m to avoid second-order light contamination, longer-wavelength data were also recorded. Following procedures in the literature, we use the G140M/F070LP configuration to extract the spectrum up to $1.8\,\mu$m, treating it as equivalent to G140M/F100LP. Since LAP1-B exhibits no features shortward of Ly$\alpha$ at $z = 6.625$ (i.e., $< 0.93\,\mu$m), contamination from second-order light is negligible in the range $1.3$--$1.8\,\mu$m.
In addition to the standard pipeline, we applied custom procedures for residual background subtraction, spatial alignment, and optimal combination of 2D spectra. These procedures are described in more detail in ref \cite{nakajima2023_jwst}.

\subsection*{Emission Line Measurements}
\label{ssec:emissionline}

We modeled emission lines using single Gaussian profiles convolved with the instrumental line spread function of each NIRSpec grating, as provided by the JWST documentation. The systemic redshift and intrinsic velocity dispersion were derived from the high S/N \Ha\ emission line and fixed for all other lines, except for \Lya, which was independently fit due to its resonant and potentially offset nature.
Line fluxes were measured by performing a chi-square minimization fit to the extracted 1D spectrum, using the associated noise spectrum to weight each data point. Flux uncertainties were estimated by summing the noise in quadrature across a $\pm$FWHM window centered on the line.
For the \CIV\ doublet, the $1548$\,\AA\ and $1550$\,\AA\ components were fit simultaneously. The total flux reported for \CIV$\lambda 1549$ corresponds to the sum of these two components.
We define a detection as a signal-to-noise ratio exceeding $3\sigma$. By this criterion, five lines are formally detected: \Ha, \Hb, \OIII$\lambda 5007$, \CIV$\lambda 1549$, and \Lya. For all other features with lower significance, we report $3\sigma$ upper limits based on the local noise.
Undetected lines include \HeII$\lambda 1640$, \CIII$\lambda\lambda 1907,1909$, \OIII$\lambda 4363$, and \OIII$\lambda 4959$. The non-detection of \OIII$\lambda 4959$ is expected given its theoretical flux ratio of $1:2.98$ relative to \OIII$\lambda 5007$, which is only marginally detected at the $3\sigma$ level. Lastly, the \OII$\lambda\lambda 3726,3729$ doublet lies outside the wavelength range covered by our observations.

The primary emission line measurements are summarized in Extended Data Table \ref{tbl:properties}.

\subsection*{Abundance Determination}
\label{ssec:abundances}

Gas-phase oxygen abundances were estimated using the \OIII$/$\Hb\ emission line ratio, one of the most widely adopted strong-line diagnostics for metallicity. In Extended Data Fig.~\ref{fig:Z_R3}, we compare this ratio against oxygen abundance using a range of calibrations based on both empirical data and theoretical models.
Empirically, metallicity indicators are often calibrated using galaxies with direct electron temperature measurements. Recent JWST observations of galaxies from $z=2$--$3$ to $9$ have enabled the construction of new empirical \OIII$/$\Hb\ calibrations down to \Oabundance\ $\simeq 7.0$ \citemeth{sanders2024, chakraborty2025}, shown as the orange and magenta curves in Extended Data Fig.~\ref{fig:Z_R3}. These high-redshift calibrations predict systematically higher \OIII$/$\Hb\ values at fixed metallicity compared to those derived from low-redshift analogs with strong Balmer emission lines \citemeth{nakajima2022_empressV}. This offset is typically interpreted as evidence for more extreme ionization conditions and/or harder ionizing spectra in high-redshift systems, though it may also reflect a selection bias toward highly ionized galaxies in these samples.
From a theoretical perspective, ref \citemeth{hirschmann2023} have derived metallicity calibrations based on cosmological simulations, which also support elevated \OIII$/$\Hb\ ratios in the low-metallicity, high-redshift regime (green curve). These are broadly consistent with predictions from photoionization models.
In our comparison, we include grids from ref \cite{NM2022} that use different stellar population assumption.
Our baseline Population II models utilize the ``Binary Population and Spectral Synthesis'' (BPASS) code (v2.2.1 \citemeth{eldridge2017,stanway2018}; gray), which assumes a continuous star formation history over 10\,Myr with a standard Kroupa IMF over $0.1$--$300$\,\Msun\ and matched stellar and gas-phase metallicities. For comparison, we also show predictions using Population III stellar models ($1$--$100$\,\Msun\ \cite{schaerer2002}; blue), with both sets of models spanning a range of ionization parameters.
A reasonable agreement is observed at low metallicities (\Oabundance\ $\simeq 7.0$) among high-redshift empirical trends, simulations, and high-ionization photoionization models with $\log U = -0.5$. Given the lack of empirical calibrations below \Oabundance\ $\simeq 7.0$, we adopt the $\log U = -0.5$ photoionization model to extend the calibration into the extremely metal-poor regime.

Our choice of a high ionization parameter ($\log U = -0.5$) for our analysis is physically motivated. There is a well-established anti-correlation between metallicity and the ionization parameter, where more metal-poor galaxies exhibit harder radiation fields and more extreme ISM conditions \citemeth{AM2013, perez-montero2014, sanders2020, nishigaki2025_dreams}. Our adopted value represents a natural extrapolation of this empirical trend to the extremely low-metallicity regime of LAP1-B.
Although $\log U = -0.5$ represents an extremely high ionization parameter, several lines of evidence favor a highly ionized interstellar medium for LAP1-B. While a direct constraint on the \OIII/\OII\ ratio is unavailable, the observed extreme Balmer line equivalent widths, EW(\Ha) $>1800$\,\AA\ and EW(\Hb) $>340$\,\AA\ ($3\sigma$), are typically associated with local extremely metal-poor galaxies exhibiting \OIII/\OII\ $\gtrsim 20$ \citemeth{nakajima2022_empressV}, corresponding to ionization parameters of at least $\log U \gtrsim -2$. This inference is consistent with the \OIII/\OII\ limit of $>3$ ($2\sigma$) in the entire LAP1 system \cite{vanzella2023_metalpoor}. Given these considerations, our adopted model serves as a reasonable and physically motivated extension of the empirical O/H-\OIII/\Hb\ relationship into the \Oabundance\ $<7.0$ regime.
We acknowledge, however, that inferred ionization parameters are model-dependent, and the intrinsic value for LAP1-B remains an open question.

At these low metallicities, the difference in the \OIII$/$\Hb\ ratio between Population II- (e.g. BPASS) and Population III-based models becomes negligible. We adopt the Population III model grid as a representative calibration for this extremely metal-deficient regime, noting that the resulting oxygen abundances would remain essentially unchanged even if a Population II grid were employed instead (see also Sect.~\ref{ssec:IMFs}). Applying this calibration to our observed \OIII$/$\Hb\ ratio of $0.69 \pm 0.28$, we derive an oxygen abundance of \Oabundance\ $= 6.31^{+0.15}_{-0.23}$.
Although an alternative explanation for the low \OIII$/$\Hb\ ratio is the presence of extremely high electron densities, exceeding the critical density of the \OIII$\lambda5007$ upper level ($n_{\rm crit} \sim 7 \times 10^5$\,cm$^{-3}$), this scenario appears unlikely for LAP1-B. At electron densities of $\sim 2 \times 10^8$\,cm$^{-3}$, collisional de-excitation would suppress \OIII\ emission, allowing a \OIII$/$\Hb\ ratio of $0.7$ to correspond to a higher oxygen abundance ($\sim 10$\% solar). However, such high densities are typically associated with dense gas in the immediate vicinity of accreting black holes, which is inconsistent with several spectroscopic features of LAP1-B. In particular, the strong hydrogen line emission relative to the continuum, evidenced by both the extreme EW(\Ha) and high ionizing photon production efficiency (\xiion), cannot be reproduced by the spectra of accreting black holes (see below). Moreover, we detect no broad component associated with the \Ha\ emission line, and thus find no spectroscopic signature of massive black hole accretion in this system.

Fig.~\ref{fig:MZR} presents a compilation of 300 galaxies at redshifts $z=4$--$12.5$ identified with JWST, serving as a comparison sample. The gas-phase metallicities of these galaxies have been derived using methods comparable to those applied to LAP1-B. This compilation incorporates results from \cite{nakajima2023_jwst, curti2024_MZR, alvarez-marquez2025, hsiao2024, hsiao2025, zavala2025, dEugenio2024}$^{,}$\citemeth{morishita2024, sarkar2025, venturi2024, marconcini2024_JD1, williams2023, schaerer2024, mowla2024, cullen2025, willott2025}. Additionally, we include the properties of present-day UFDs as a local reference; the details regarding the selection and derivation of these UFD data points are provided in Sect.~\ref{ssec:UFDs}.
\\

A second notable metal line detected in the spectrum of LAP1-B is \CIV$\lambda 1549$.
The simultaneous detection of both \CIV$\lambda 1549$ and \OIII$\lambda 5007$ is significant, as these are the dominant cooling lines in ionized gas in extremely metal-poor environments \cite{NM2022}. Their relative strength allows us to estimate the carbon-to-oxygen abundance ratio (C/O), which provides key insights into the chemical enrichment history of the system.
To derive the ionic abundance ratio C$^{3+}$/O$^{2+}$, we employ the latest atomic data implemented in the \texttt{PyNeb} package. The conversion between the observed \CIV/\OIII\ line flux ratio and the ionic abundance ratio is given by: 
\begin{align}
\log (\mathrm{C}^{3+}/\mathrm{O}^{2+}) = 
	\log\left( \frac{\mathrm{C}\,\mbox{\sc iv}\lambda 1549}{[\mathrm{O}\,\mbox{\sc iii}]\lambda 5007}\right) 
	- 1.736 + \frac{2.758}{t_\mathrm{e}} + 0.004 t_\mathrm{e} - 0.083\log(t_\mathrm{e})
	\label{eq:co}
\end{align}
where $t_{\rm e}$ is the \OIII\ electron temperature in units of $10^4$\,K. This expression is valid over a wide range of temperatures ($0.4 < t_{\rm e} < 5.0$) 
and assumes an electron density of $1,000$\,cm$^{-3}$. We note that this relation remains nearly unchanged for electron densities up to $\sim10^5$\,cm$^{-3}$, i.e., below the critical density for collisional de-excitation of the \OIII$\lambda5007$ line. This expression is consistent with previous formulations based on UV lines \citemeth{perez-montero2017}.

To ensure an accurate measurement, we first assess the potential impact of dust extinction. The observed Balmer decrement, \Ha/\Hb\ $= 2.83 \pm 0.84$, is consistent with the intrinsic case B ratio, indicating negligible internal reddening. We therefore adopt the observed \CIV/\OIII\ ratio directly as the intrinsic value for the C$^{3+}$/O$^{2+}$ calculation.
A direct measurement of the electron temperature is not possible, as the temperature-sensitive auroral line \OIII$\lambda4363$ is undetected. The current upper limit on the \OIII$\lambda4363$/\OIII$\lambda5007$ ratio ($<0.85$ at $3\sigma$) does not provide a meaningful constraint on $t_{\rm e}$. Instead, 
we estimate the electron temperature using \texttt{PyNeb}, based on the observed \OIII/\Hb\ ratio and the inferred oxygen abundance. As discussed above, the high ionization parameter in LAP1-B implies that singly ionized oxygen contributes minimally to the total oxygen budget, justifying the use of \OIII\ as representative of total oxygen. This yields an estimated electron temperature of $t_{\rm e} = 2.5$--$2.6$ ($10^4$\,K). 
By propagating the uncertainties in both the \CIV/\OIII\ flux ratio and the estimated electron temperature, we derive an ionic abundance ratio of $\log(\mathrm{C}^{3+}/\mathrm{O}^{2+}) = -0.07^{+0.16}_{-0.24}$. Since the \CIII$\lambda1909$ line is not detected, we assume that C$^{3+}$/O$^{2+}$ approximates the total C/O abundance ratio. This is a conservative assumption, as any additional contribution from C$^{2+}$ would increase the total C/O ratio, further strengthening the case for carbon enhancement in this system.

The resulting C/O abundance in LAP1-B corresponds to approximately $1$--$2$ times the solar value. Combined with its extremely low oxygen abundance, this places LAP1-B in a notably extreme region of the C/O versus O/H diagram (Fig.~\ref{fig:co_oh}).
This figure compares our measurement to a comprehensive sample of other astrophysical objects, including high-redshift galaxies with carbon measurements obtained through JWST spectroscopy \cite{ji2024, dEugenio2024, naidu2026_z14p44}$^{,}$\citemeth{arellano-cordova2022_ero, jones2023, isobe2023_cno, stiavelli2023, castellano2024_z12, schaerer2024, topping2024, hsiao2025_carbon, curti2025, carniani2025}, damped Lyman-$\alpha$ systems \citemeth{cooke2017}, nearby \HII-regions and star-forming galaxies, galaxies at $z=1.5$--$3.5$ (as compiled in ref \citemeth{jones2023}), as well as individual stars in the Milky Way \cite{akerman2004} \citemeth{gustafsson1999, spite2005, bensby2006, fabbian2009, nissen2014}. 
For a direct comparison with local fossils, we plot individual stars identified in present-day UFDs; the selection and characterization of these stellar samples are detailed in Sect.~\ref{ssec:UFDs}. 
The position of LAP1-B is broadly consistent with theoretical predictions for enrichment by metal-deficient Population III stars.

An alternative to consider is enrichment from Asymptotic Giant Branch (AGB) stars. This is plausible, as chemical evolution proceeds differently in low-mass systems than in Milky Way-like models, potentially allowing for AGB-driven C/O enhancement to occur at a lower metallicity. However, this hypothesis is inconsistent with other chemical evidence from LAP1-B's likely local descendants, the UFDs. AGB nucleosynthesis produces s-process elements, such as Barium (Ba), alongside carbon. If AGB stars were the primary source of enrichment, UFDs should show significant Barium enhancement. Instead, observations consistently reveal very low [Ba/Fe] ratios across the UFD population \citemeth{frebel2014, tsujimoto2014}, reinforcing the conclusion that the chemistry of LAP1-B is likely a fossil record of enrichment from massive, metal-deficient stars.
\\

Further spectroscopic characterization, including an assessment of the ionizing radiation (Extended Data Fig.~\ref{fig:diagnostics_popIII}), is presented in the Supplementary Information.

\subsection*{Stellar, Gas, and Dynamical Mass}
\label{ssec:masses}

The stellar, gas, and dynamical masses of LAP1-B are estimated using a combination of photometric constraints, star-formation scaling relations, and kinematic analysis. Detailed methodologies and the treatment of gravitational lensing magnification are provided in the Supplementary Information.

\subsection*{Comparison with Present-day UFDs}
\label{ssec:UFDs}

The methodologies for comparing the high-redshift properties of LAP1-B with the stellar populations of local UFD relics, including the mass–metallicity relation, carbon-to-oxygen abundance trends, and the conversion of stellar [Fe/H] measurements to oxygen abundance (Extended Data Fig. \ref{fig:co_oh_MWhalostars}), are fully detailed in the Supplementary Information.

\subsection*{Photoionization model calculations}
\label{ssec:photoionizationmodels}

The photoionization model predictions shown in Fig.~\ref{fig:c4o3_o3hb} as well as in Extended Data Figs.~\ref{fig:Z_R3}, \ref{fig:diagnostics_popIII} \ref{fig:models_popIII_topheavyIMF} and \ref{fig:models_popIII_bottomheavyIMF} are based on the grid developed in ref \cite{NM2022} using \texttt{Cloudy} (version 13.05; \citemeth{ferland1998,ferland2013}).
In the original models, the carbon-to-oxygen abundance ratio (C/O) scales with metallicity following the empirical relation \citemeth{dopita2006}, in which $\log$~(C/O) $\simeq -0.9$ (corresponding to $\sim$0.2$\times$ solar) at low metallicities (\Oabundance\ $\lesssim 7$), increasing quadratically toward the solar value at solar metallicity.
For the chemically enriched Population II models, we incorporate dust physics and element-specific depletion factors to account for the removal of metals from the gas phase. We adopt the default depletion from \texttt{Cloudy} ($0.4$ for carbon and $0.6$ for oxygen \citemeth{ferland2013}; see also ref \citemeth{jenkins2009}), which represent the fraction of an element remaining in the gas-phase rather than being locked in grains. We further assume that the dust-to-gas ratio scales linearly with the total metallicity.

To specifically investigate the unique properties of LAP1-B, we extended this framework in three significant ways. 
First, to explore the impact of its high carbon abundance, we generated new model grids with fixed, enhanced C/O ratios of 1$\times$ and 2$\times$ the solar value, independent of metallicity. These models, a subset of which were also presented in ref \cite{dEugenio2024}, are shown in Fig.~\ref{fig:c4o3_o3hb}, allowing us to properly interpret the diagnostic diagrams for a carbon-rich environment. 
Second, to test the nature of the ionizing source, we performed an additional suite of calculations exploring the impact of the IMF. This new analysis considers a range of IMFs for both zero-metal (Population III) and chemically-enriched (Population II) stellar populations, including more extreme cases than in the original models. This comprehensive modeling is detailed in the Supplementary Information (Extended Data Figs.~\ref{fig:models_popIII_topheavyIMF} and \ref{fig:models_popIII_bottomheavyIMF}).
Third, to evaluate the impact of assumed dust physics, we performed additional Population II calculations that exclude dust grains from the photoionized regions and assume zero depletion of elements. This modification affects both the gas-phase chemical composition and the thermal balance of the cloud, specifically removing the contribution of photoelectric heating by dust \citemeth{vanHoof2004}.
For all photoionization models, a fiducial electron density of $1,000$\,cm$^{-3}$ is adopted, consistent with the assumptions in the original model suite \cite{NM2022}.

\clearpage

\section*{Extended Data}
\label{sec:extended_data}

\newcounter{extfigure}
\renewcommand{\thefigure}{\arabic{extfigure}}
\renewcommand{\figurename}{Extended Data Fig.}
\renewcommand{\theHfigure}{ext.\arabic{extfigure}}
\setcounter{figure}{0}

\newcounter{exttable}
\renewcommand{\thetable}{\arabic{exttable}}
\renewcommand{\tablename}{Extended Data Table}
\renewcommand{\theHtable}{ext.\arabic{exttable}}
\setcounter{table}{0}

\stepcounter{extfigure} 
\begin{figure}[h!]
    \begin{center}
    \includegraphics[bb=0 0 533 510, width=0.75\columnwidth]{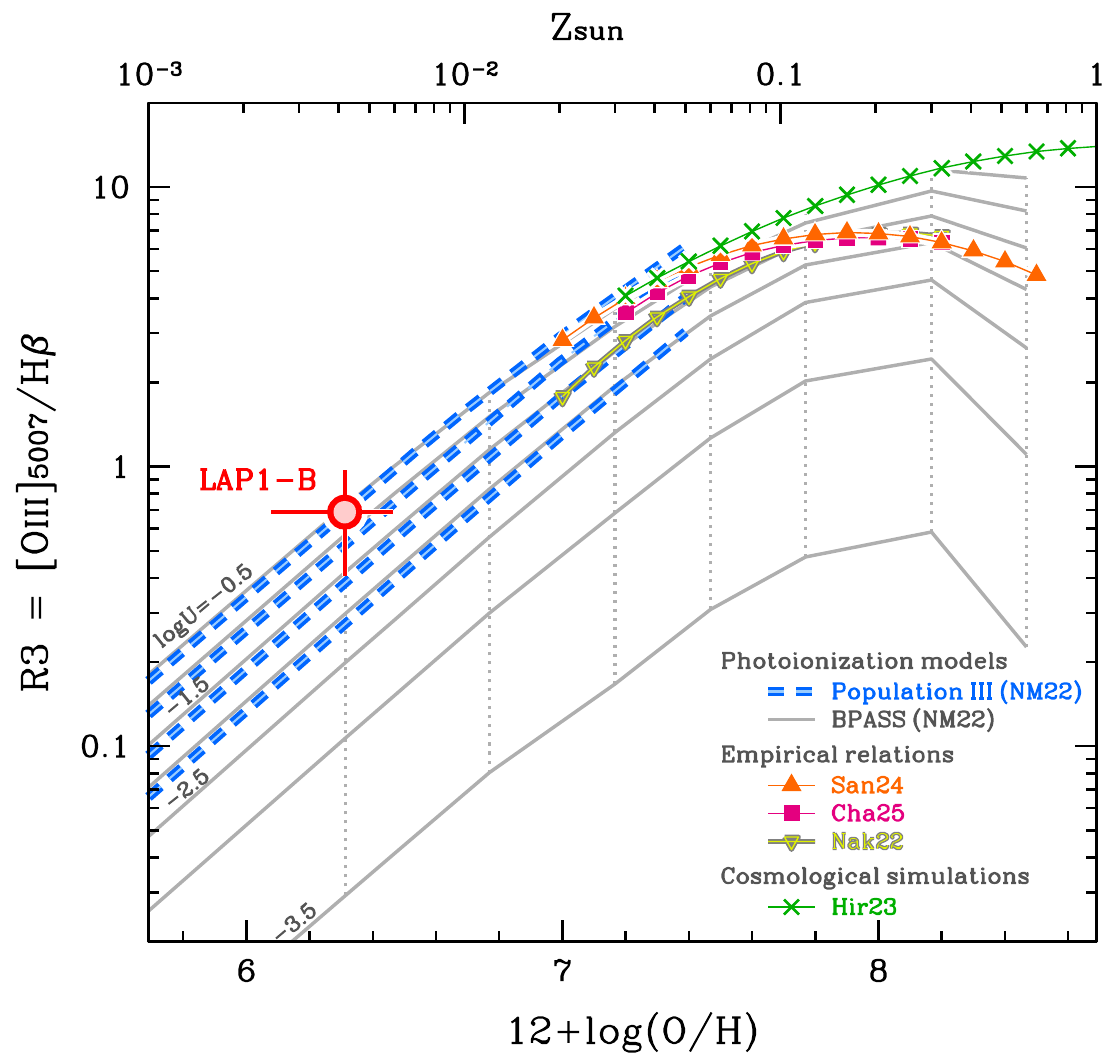}
    \caption{%
	\textbf{Oxygen abundance diagnostics based on \OIII$\lambda 5007$/\Hb}.
	The relationship between the \OIII$\lambda 5007$/\Hb\ line ratio and gas-phase oxygen abundance is shown. Curves with symbols represent empirically calibrated relations derived from galaxies at $z=2-9$ \protect\citemeth{sanders2024, chakraborty2025}, extremely metal-poor galaxies at $z\simeq 0$ with EW(\Hb) $\geq 200$\,\AA\ \protect\citemeth{nakajima2022_empressV}, and cosmological simulations \protect\citemeth{hirschmann2023}, as indicated in the legend. Each curve is plotted over the range explored in the respective studies. 
	Overlaid are theoretical predictions from photoionization models \cite{NM2022}, using two types of stellar ionizing sources: chemically enriched BPASS stellar populations (continuous star formation history over 10\,Myr with a Kroupa IMF up to 300\,\Msun) and zero-metallicity Population III stars (Salpeter IMF, 1-100\,\Msun\ \cite{schaerer2002}). Different curves correspond to varying ionization parameters: from $\log U = -0.5$ to $-2.0$ for the Population III models, and extending down to $\log U = -3.5$ for the BPASS models.
	The Population III model with the highest ionization parameter is adopted to derive the oxygen abundance of LAP1-B (see text for justification), whose observed \OIII/\Hb\ ratio and derived oxygen abundance are indicated in red.
	}
    \label{fig:Z_R3}
    \end{center} 
\end{figure}

\stepcounter{extfigure} 
\begin{figure}[h!]
  \centering
    \begin{tabular}{c}
      \begin{minipage}{0.49\hsize}
        \begin{center}
          \includegraphics[bb=0 0 541 470, width=1.\columnwidth]{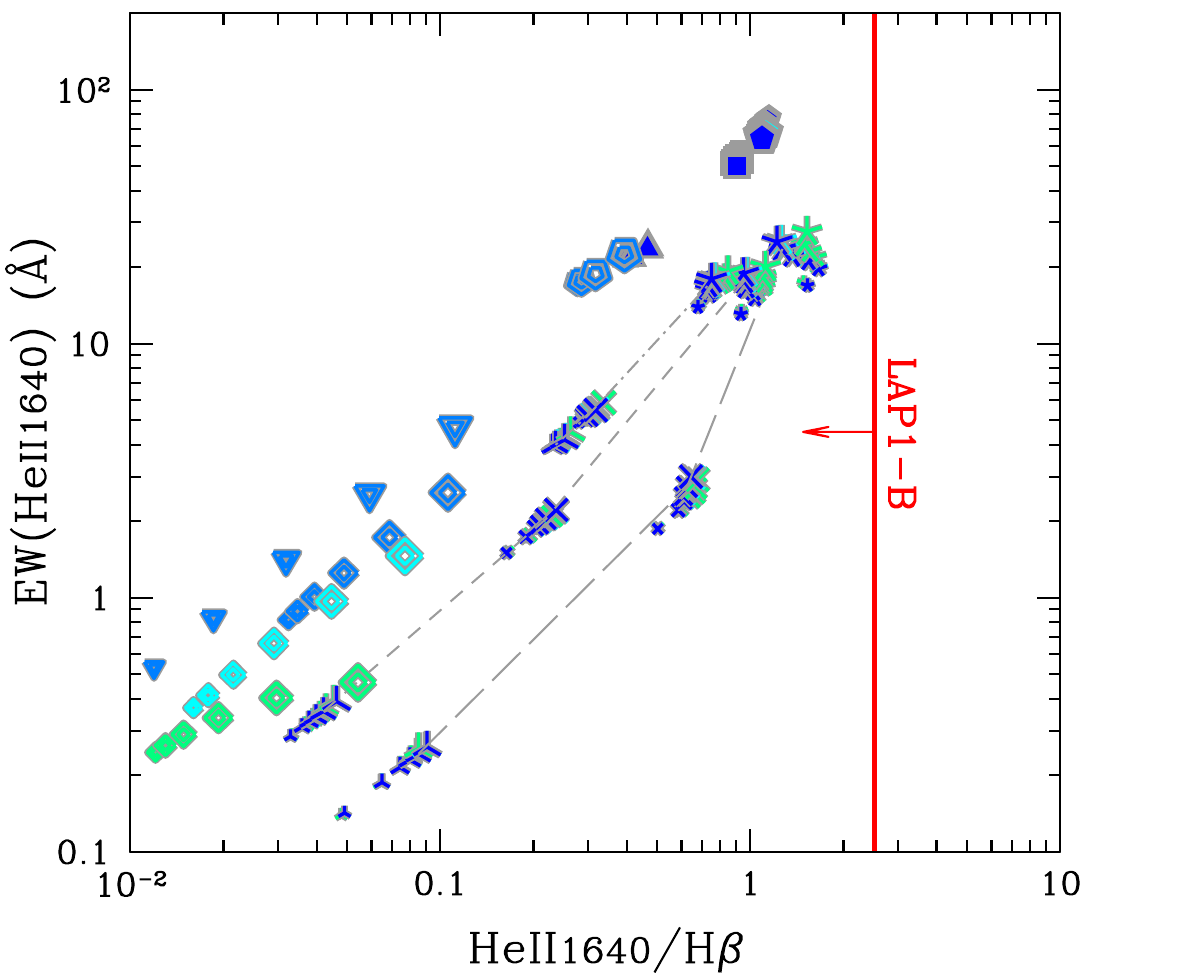}
        \end{center}
      \end{minipage}
      \begin{minipage}{0.49\hsize}
        \begin{center}
         \includegraphics[bb=0 0 546 470, width=1.\columnwidth]{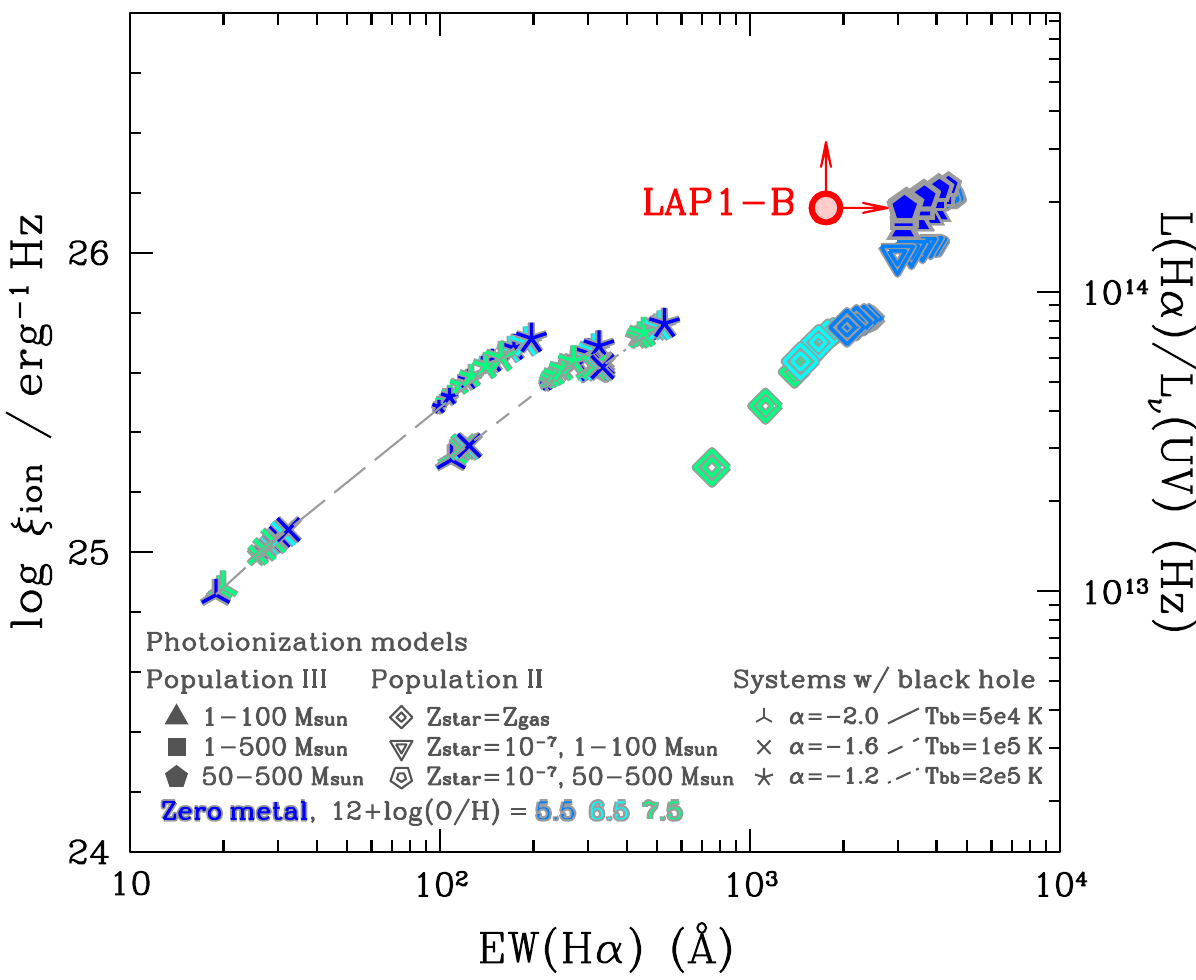}
        \end{center}
      \end{minipage}
    \end{tabular}
    \caption{%
	\textbf{Diagnostic diagrams probing the shape of ionizing spectrum.}
	LAP1-B (red) is compared with photoionization model predictions \cite{NM2022} and some from extremely metal-poor models from this work using two diagnostic plots: EW(\HeII$\lambda 1640$) versus \HeII/\Hb\ line ratio (left), and the ionizing photon production efficiency (\xiion) versus EW(\Ha) (right). \xiion\ is derived as the \Ha\ luminosity divided by the UV continuum flux density, with the resulting values plotted on the right-hand y-axis. The measurement for LAP1-B conservatively assumes a zero escape fraction of ionizing photons; this provides a lower limit, as a non-zero escape fraction would imply an even higher intrinsic \xiion.
	Model predictions shown in both panels correspond to different ionizing sources: Population III stars, chemically enriched Population II stars, and accreting black holes (see ref \cite{NM2022} for details). Symbol shapes distinguish between these ionizing sources, while colors indicate different gas-phase metallicities, as in the legend. Note that the newly developed $Z=10^{-7}$ Population II models with a $50$--$500$\,\Msun\ IMF overlap the Population III region in the right-hand \xiion\ panel, offering an alternative explanation for the observed value.
	}
    \label{fig:diagnostics_popIII}
\end{figure}

\stepcounter{extfigure} 
\begin{figure}[h!]
    \begin{center}
    \includegraphics[bb=0 0 521 354, width=0.75\columnwidth]{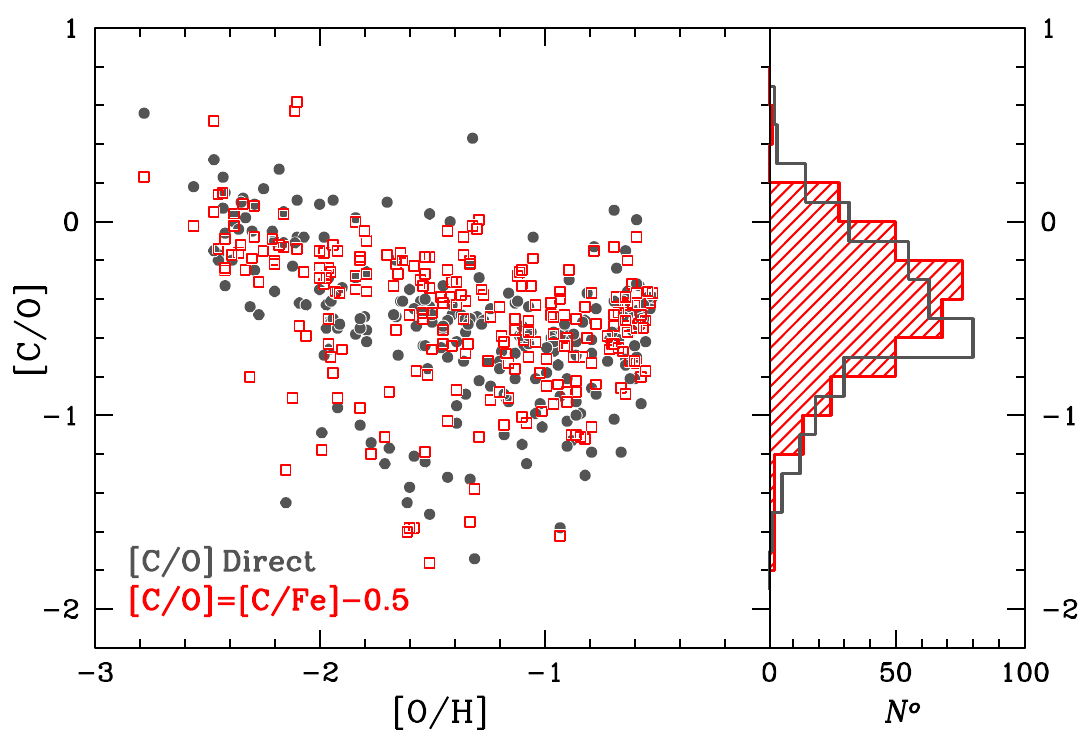}
    \caption{%
	\textbf{Chemical abundance patterns observed in Milky Way halo stars.}
	(Main panel:) The [C/O] vs.~[O/H] abundance plane for a sample of metal-poor stars from the Milky Way halo, drawn from the SAGA database. The selected stars are those with individual abundance measurements for C, O, and Fe. Gray points represent the directly measured [C/O] ratios. Red points show the inferred [C/O] ratios for the same stars, calculated by assuming a constant [O/Fe] $=+0.5$ (i.e. [C/O] = [C/Fe] $-0.5$). 
	(Right sub-panel:) The distributions of the measured (gray) and inferred (red) [C/O] ratios. The excellent agreement between the two distributions validates our use of [O/Fe] $=+0.5$ as a representative value.
	}
    \label{fig:co_oh_MWhalostars}
    \end{center} 
\end{figure}

\stepcounter{extfigure} 
\begin{figure}[h!]
  \centering
    \begin{tabular}{c}
      \begin{minipage}[b]{0.49\hsize}
        \begin{center}
          \includegraphics[bb=0 0 532 470, width=.9\columnwidth]{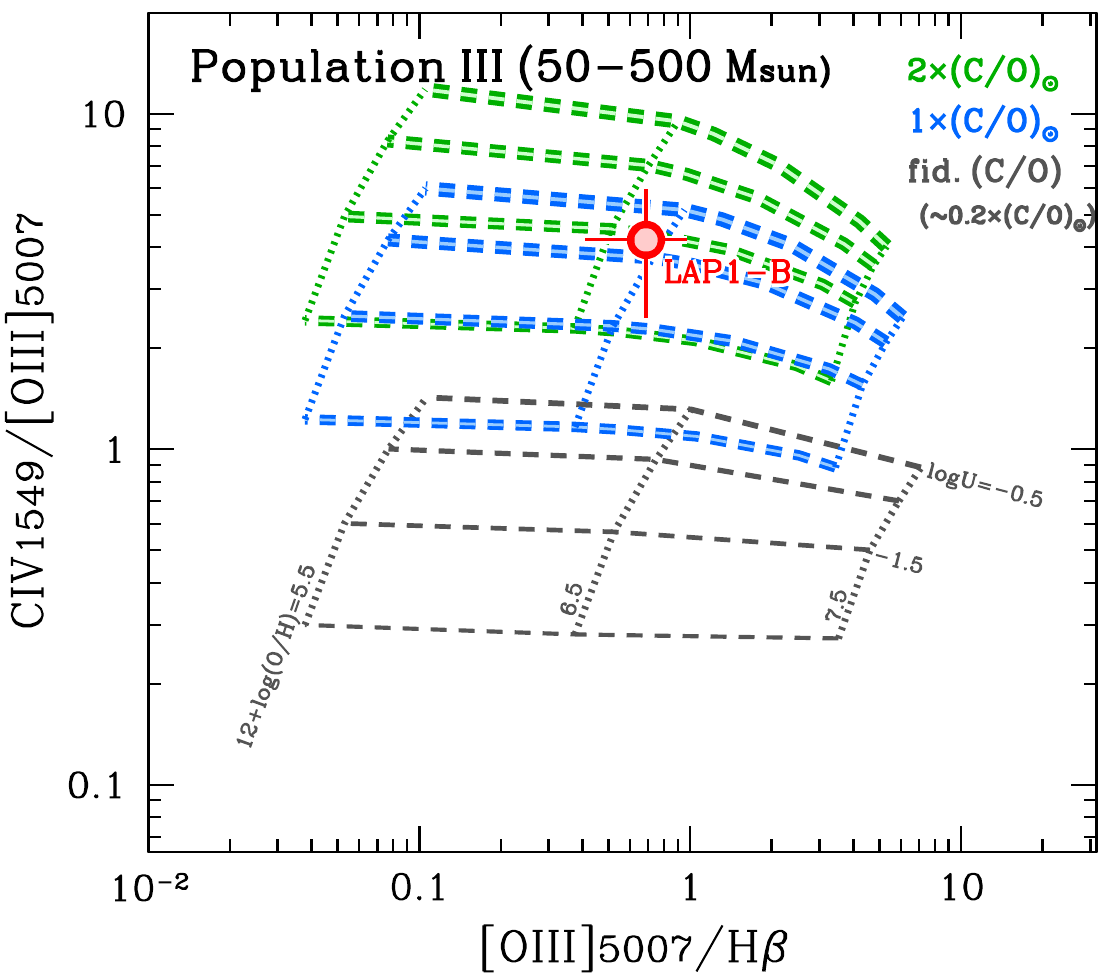}
        \end{center}
      \end{minipage}
      \begin{minipage}[b]{0.49\hsize}
        \begin{center}
         \includegraphics[bb=0 0 533 510, width=.9\columnwidth]{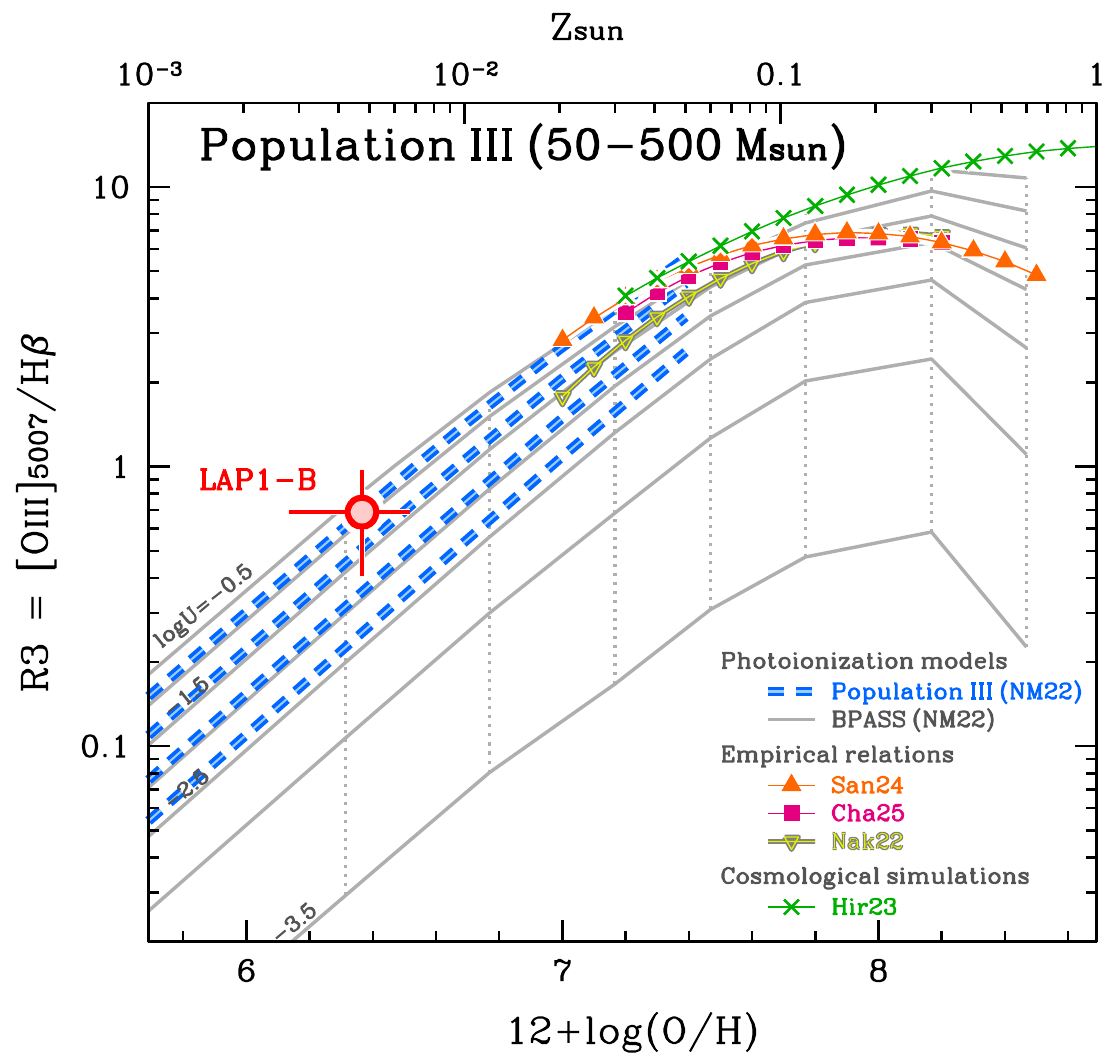}
        \end{center}
      \end{minipage}
    \end{tabular}
    \caption{%
	\textbf{
	Same as Fig.~\ref{fig:c4o3_o3hb} (left) and Extended Data Fig.~\ref{fig:Z_R3} (right) but the Population III models are replaced with the top-heavy IMF of $50$--$500$\,\Msun. }
	}
    \label{fig:models_popIII_topheavyIMF}
\end{figure}

\stepcounter{extfigure} 
\begin{figure}[h!]
  \centering
    \begin{tabular}{c}
      \begin{minipage}[b]{0.49\hsize}
        \begin{center}
          \includegraphics[bb=0 0 532 470, width=.9\columnwidth]{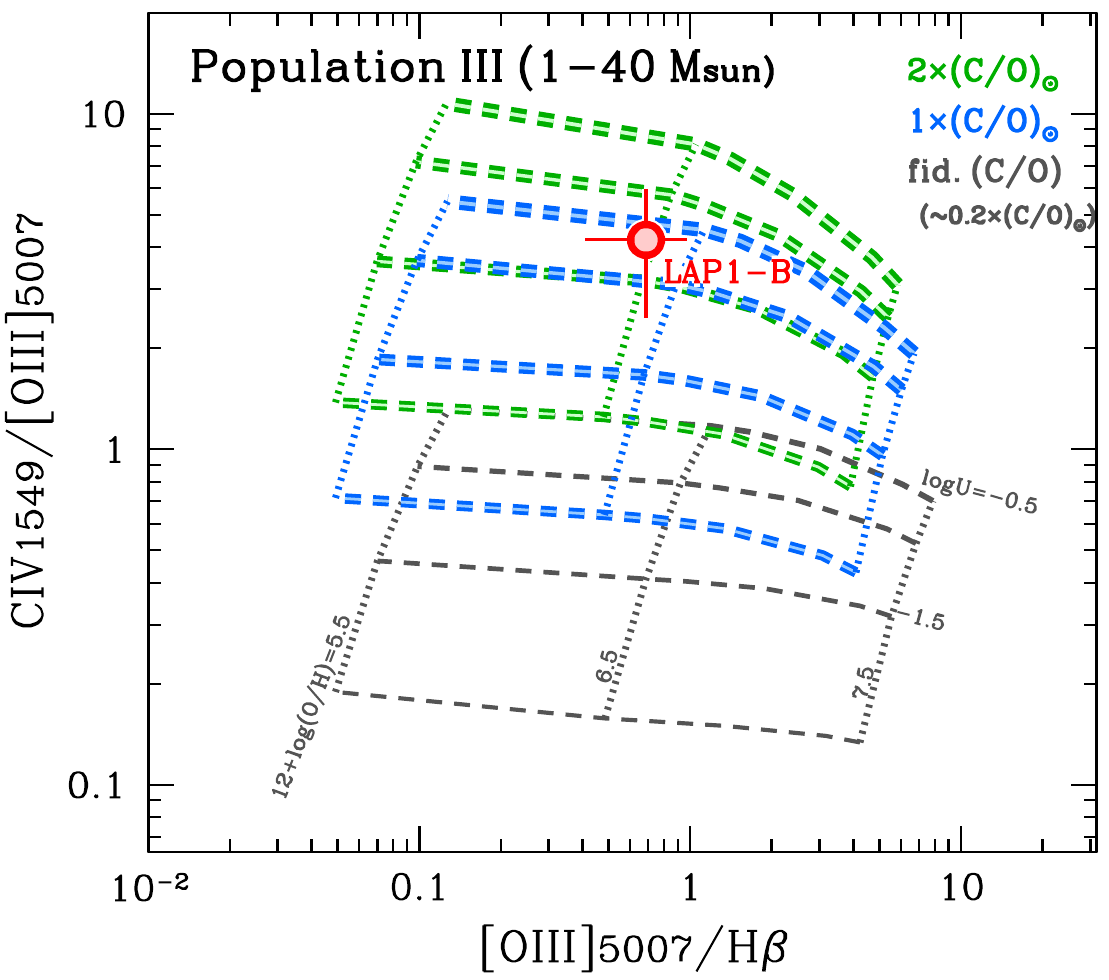}
        \end{center}
      \end{minipage}
      \begin{minipage}[b]{0.49\hsize}
        \begin{center}
         \includegraphics[bb=0 0 533 510, width=.9\columnwidth]{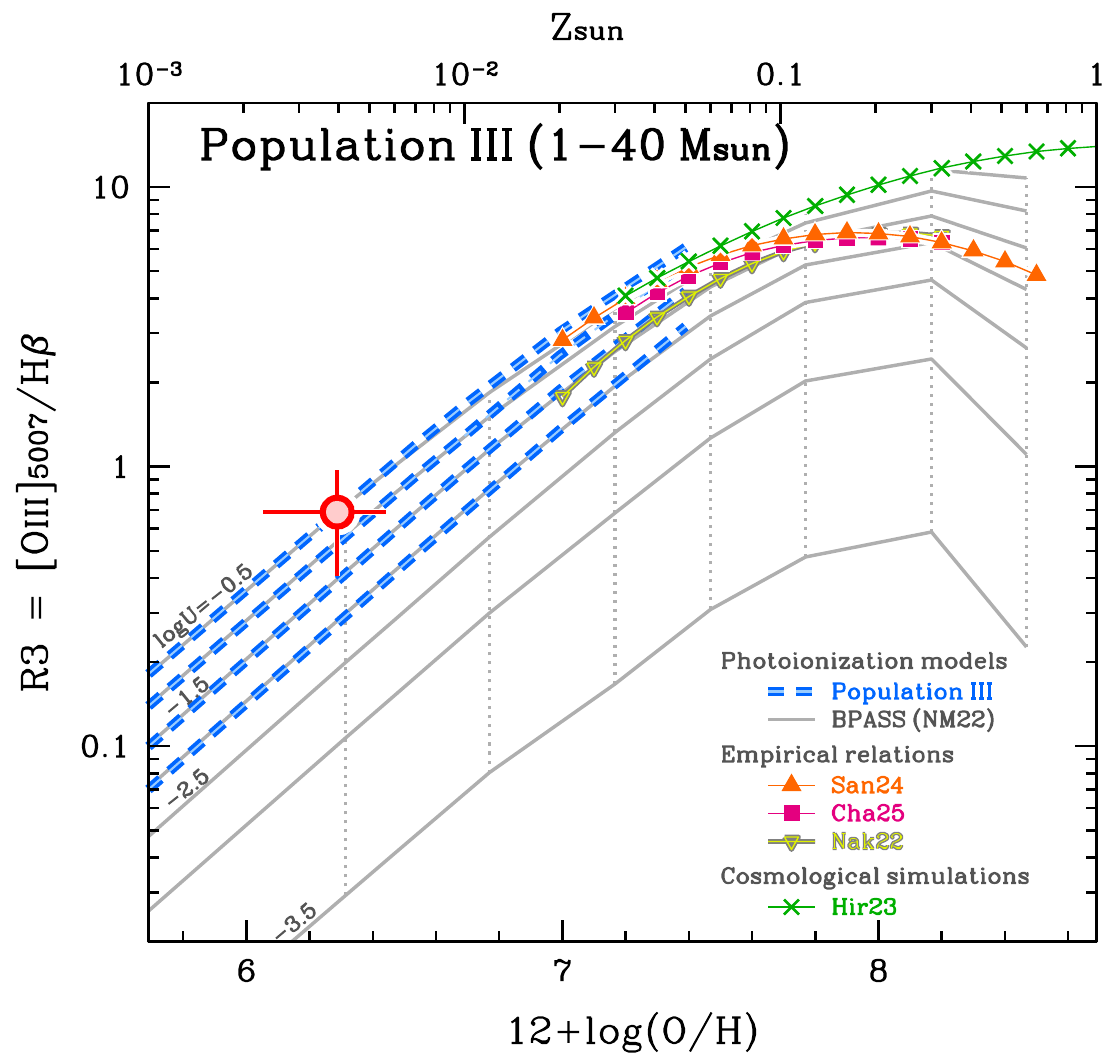}
        \end{center}
      \end{minipage}
    \end{tabular}
    \caption{%
	\textbf{
	Same as Fig.~\ref{fig:c4o3_o3hb} (left) and Extended Data Fig.~\ref{fig:Z_R3} (right) but the Population III models are replaced with the top-heavy IMF of $1$--$40$\,\Msun. }
	}
    \label{fig:models_popIII_bottomheavyIMF}
\end{figure}

\stepcounter{exttable} 
\begin{table}[h!]
\caption{Measured emission-line fluxes and upper limits for LAP1-B}
\label{tbl:properties}%
\begin{tabular}{ll}
\toprule
Line & Flux \\
 & ($10^{-19}$\,\ergscm) \\
\midrule
\Lya\ & $6.08 \pm 1.70$ \\
\CIV$\lambda\lambda 1548,1550$ & $2.12 \pm 0.61$ \\
\HeII$\lambda 1640$ & $<1.85$ \\
\CIII$\lambda\lambda 1907,1909$ & $<4.55$ \\
\OII$\lambda\lambda  3726, 3729$ & $\cdots$ \\
\OIII$\lambda 4363$ & $<0.43$ \\
\Hb\ & $0.73 \pm 0.20$ \\
\OIII$\lambda 5007$ & $0.50 \pm 0.15$ \\
\Ha\ & $2.07 \pm 0.25$ \\
\botrule
\end{tabular}
\footnotetext{Fluxes are uncorrected for lensing magnification. Upper limits correspond to $3\sigma$ significance.}
\end{table}

\clearpage

\section*{Data availability}
\label{sec:data_availability}

The reduced JWST/NIRSpec spectroscopic data supporting the findings of this study will be made publicly available via Zenodo (URL to be provided upon publication).
The corresponding raw data under Program ID \#4750 (PI: K.~Nakajima) are publicly accessible through the STScI MAST archive: \url{https://mast.stsci.edu/portal/Mashup/Clients/Mast/Portal.html}.

\section*{Code availability}
\label{sec:code_availability}

The data were reduced using the JWST calibration pipeline (version 1.17.1) with the CRDS context jwst\_1298.pmap, both provided by STScI. The pipeline is publicly available at: \url{https://github.com/spacetelescope/jwst}.
Additional analysis made use of the publicly available software packages:
\texttt{AstroPy} \citemeth{astropy2013,astropy2018,astropy2022},
\texttt{Cloudy} \citemeth{ferland1998,ferland2013}, and
\texttt{PyNeb} \citemeth{luridiana2015}.


\bibliographystylemeth{sn-mathphys-num}
\bibliographymeth{Refs_sn-article}{}


\section*{Acknowledgements}
\label{sec:acknowledgements}

We would like to thank M. Leveille and D. Karakla for their tremendous support during the preparation of the observations. 
We are grateful to R. Maiolino, A. Fialkov, and T. Gessey-Jones for insightful discussions and for generously providing access to their developmental Population III synthesis code \citemeth{gessey-jones2022}, and to D. Schaerer for providing his extremely metal-poor stellar population synthesis models \cite{schaerer2003}. Both theoretical models were essential for this analysis. We also thank N. Tominaga for helpful discussions regarding chemical enrichment from Population III stars. Finally, we thank the anonymous referees for their constructive comments, which sharpened our arguments and significantly improved the presentation of this work.
This work is based on observations made with the NASA/ESA/CSA James Webb Space Telescope. The data were obtained from the Mikulski Archive for Space Telescopes at the Space Telescope Science Institute, which is operated by the Association of Universities for Research in Astronomy, Inc., under NASA contract NAS 5-03127 for JWST. These observations are associated with program \#4750 (NIRSpec) as well as \#1176 (GTO, NIRCam) and \#1208 (GTO, NIRCam).
This paper is supported by World Premier International Research Center Initiative (WPI Initiative), MEXT, Japan, as well as the joint research program of the Institute of Cosmic Ray Research (ICRR), the University of Tokyo. 
In addition, we acknowledge support from JSPS KAKENHI Grant: JP20K22373 and JP24K07102 (K.N.), JP20H00180, JP21H04467, and JP25H00674 (M.O.), JP24H00245 (Y.H.), JP24KJ0202 (Y.I.), and JP23H00132 (T.T.). 
E.V. acknowledges financial support through grants INAF GO Grant 2022 ``The revolution is around the corner: JWST will probe globular cluster precursors and Population III stellar clusters at cosmic dawn'' and INAF GO Grant 2024 ``Mapping Star Cluster Feedback in a Galaxy 450 Myr after the Big Bang'', and by the European Union - NextGenerationEU within PRIN 2022 project n.20229YBSAN - Globular clusters in cosmological simulations and lensed fields: from their birth to the present epoch.
M.N. is supported by JST, the establishment of university fellowships towards the creation of science technology innovation, Grant Number JPMJFS2136.

\section*{Author contributions}
\label{sec:author_contributions}

K.N. was the Principal Investigator of the JWST/NIRSpec program and led the observing proposal, planning, and data reduction. K.N. performed the spectroscopic analysis, led the core interpretation of the data, developed the underlying theoretical framework, and wrote the manuscript.
M.O. developed the observational strategy, initiated the core interpretation of LAP1-B's evolutionary connection to present-day ultra-faint dwarfs, and facilitated the interdisciplinary discussions that shaped the paper.
Y.H. performed the NIRCam data reduction and imaging analysis, which were critical for the stellar mass estimation and observation planning.
E.V. conducted the gravitational lensing analysis and utilized the lens model to measure the spatial extent of LAP1-B.
T.T. provided critical insights into the connection between LAP1-B and present-day ultra-faint dwarfs, specifically regarding the comparison of oxygen and iron abundances, and contributed to the discussions on the chemical enrichment processes responsible for the elevated carbon-to-oxygen ratio.
M.O., E.V., Y.H., Y.O., and M.N. contributed to the observational design and proposal preparation.
M.O., Y.H., Y.O., Y.I., Y.X., H.U., and Y.Z. provided scientific input to the original proposal and the final manuscript. 
F.N. contributed to the scientific discussion regarding massive star formation.
All authors contributed to the discussion of results and the final editing of the manuscript.

\section*{Competing interests}
\label{sec:competing_interests}

The authors declare no competing interests.

\clearpage

\section*{Supplementary Information}
\label{sec:supplementary}

\subsection*{Other Key Spectroscopic Features}
\label{ssec:otherspec}

To assess the nature of star formation and explore the potential signatures of Population III stellar populations in LAP1-B, we derive constraints on several key quantities: the \HeII$\lambda 1640$ emission line, the equivalent widths (EWs) of Balmer lines, the ionizing photon production efficiency (\xiion), and \Lya\ emission.
\\

We first examine the \HeII$\lambda 1640$ line, a critical diagnostic of extremely metal-poor or metal-free stellar populations. Adopting the systemic redshift of $z = 6.625\pm 0.001$ as determined by \Ha, we do not identify a signal at the expected wavelength at the $3\sigma$ level. While no constraint on the EW is obtained due to the lack of the continuum flux density detection in the rest-frame ultraviolet using NIRCam, the line flux ratio is constrained to be \HeII/\Hb\ $<2.5$ ($3\sigma$). As illustrated in the left panel of Extended Data Fig.~\ref{fig:diagnostics_popIII}, this upper limit is not a strong constraint. It is consistent with the predictions for both Population III stars and Population II stellar populations explored in this work \cite{NM2022}. Therefore, this diagnostic alone cannot distinguish between these two scenarios.
\\

An additional open question is the nature of the ionizing source, specifically, whether it arises from stellar populations with or without metallicity, or from non-thermal processes such as metal-poor black holes or low-luminosity AGNs. 
To distinguish between these scenarios, we investigate two continuum-normalized observables: the rest-frame equivalent width of \Ha, and the ionizing photon production efficiency, \xiion, which together probe the shape of the radiation field from the UV-optical regime (a few electron volts) up to energies exceeding $13.6$\,eV.
These quantities are compared in the right panel of Extended Data Fig.~\ref{fig:diagnostics_popIII}, where model tracks for different source types are overlaid. Based on the non-detection of the rest-frame optical stellar continuum, we derive a $3\sigma$ lower limit of EW(\Ha)~$>1800$\,\AA\ and EW(\Hb)~$>340$\,\AA. Adopting a stricter $5\sigma$ criterion yields EW(\Ha)$>880$\,\AA\ and EW(\Hb)~$>170$\,\AA.
The ionizing photon production efficiency, \xiion, is defined as the ratio between the rate of hydrogen-ionizing photons ($Q_{\mathrm{H}^0}$) and the non-ionizing UV luminosity ($L_{\mathrm{UV}}$), i.e., \xiion\ $=Q_{\mathrm{H}^0}/L_\mathrm{UV}$.
The production rate of hydrogen-ionizing photons is governed by the presence of young, massive stars and is best constrained by hydrogen recombination lines, specifically \Ha\ in our case. We adopt the conversion relation \citesupp{LH1995} to estimate $Q_{\mathrm{H}^0}$ from the observed \Ha\ luminosity. This relation assumes that all ionizing photons are absorbed within the nebula and reprocessed into recombination lines, corresponding to an escape fraction $f_{\mathrm{esc}} = 0$. 
The upper limit on the UV luminosity, $L_{\mathrm{UV}}$, is estimated from a stack of deep JWST/NIRCam photometry bands where the continuum remains undetected. This allows us to derive a lower limit on the ionizing photon production efficiency of $\log$~(\xiion/\ergHz) $> 26.1$ at the $3\sigma$ confidence level, and $>25.9$ at $5\sigma$.
We note that the derivations of equivalent widths and \xiion\ presented above implicitly assume identical magnification for both the nebular emission and the stellar continuum. This assumption is commonly adopted in the literature. However, if the nebular line emission and the stellar continuum emission arise from regions of different sizes, they may be subject to differential magnification. In such a case, the observed equivalent widths and \xiion\ would be amplified.
Since we currently lack a clear characterization of the stellar continuum morphology, we do not explore this possibility further and instead rely on the widely used assumption that magnification is not wavelength- or component-dependent. This remains an open question that warrants detailed lens modeling in future work.

Before interpreting the nature of the ionizing source, we first establish that LAP1-B is a self-gravitating system with its own internal power source, rather than a passive gas cloud externally illuminated by a nearby source like LAP1-A. The modest intrinsic velocity dispersion of the gas ($\sigma\simeq 58$\,\kms) within such a compact region is consistent with motions inside a deep gravitational potential well. This strongly supports the interpretation that LAP1-B is a self-contained, star-forming system.

Having established this, we can constrain the nature of its internal power source. 
The right panel of Extended Data Fig.~\ref{fig:diagnostics_popIII} compares our constraint with photoionization model predictions \cite{NM2022}, including both stellar and black hole-driven ionizing sources. These models assume a maximally young stellar population, yielding the highest possible \xiion\ and equivalent width of Balmer lines (e.g. \Ha) for a given metallicity and ionization parameter. In practice, older stellar populations would increase the non-ionizing continuum relative to the ionizing photon output, thereby lowering both \xiion\ and the Balmer line equivalent widths. 
As shown in Fig.~\ref{fig:diagnostics_popIII} (right), the observed position of LAP1-B falls well above the predictions for black hole-dominated sources and standard Population II stellar populations. Instead, it is compatible only with two exotic scenarios: Population III stellar population, or an extremely metal-deficient ($Z\sim 10^{-7}$) Population II model with an IMF composed exclusively of very massive stars ($50$--$500$\,\Msun; see Sect.~\ref{ssec:IMFs}). This conclusion is robust; a non-zero escape fraction of ionizing photons would only increase the inferred intrinsic \xiion, strengthening the case against conventional models.
Similar implications regarding the presence of very massive stars in metal-poor environments are discussed in \cite{vanzella2024_T2c, schaerer2025}$^{,}$\citesupp{martins2025}.

The observed lower limit on \xiion\ places a notably stringent constraint on the stellar age, approaching the theoretical maximum for stellar emission. For context, the maximum \xiion\ predicted for a zero-age Population III starburst is $\log$(\xiion) $\approx 26.2$. Under a constant star formation scenario, \xiion\ decreases with age as the non-ionizing UV continuum accumulates relative to the short-lived massive stars producing the ionizing photons. We find that even models assuming a Population III IMF composed exclusively of very massive stars ($50$--$500$\,\Msun) predict \xiion\ values that drop below our observed $3\sigma$ lower limit after only $\sim 5$--$10$\,Myr. Consequently, to be consistent with these theoretical limits, the star-forming population in LAP1-B must be extremely young, with an age of $\lesssim 5$\,Myr. This extremely young age is fully consistent with the stringent upper limit on the stellar mass, as the system has not yet had time to accumulate a significant evolved stellar population.

Collectively, these findings supports the interpretation that LAP1-B is observed in a brief but critical stage of its assembly. In this phase, the first massive stars have likely already exploded as supernovae, modestly enriching the interstellar medium, while the ionizing flux is provided by a co-eval or subsequent generation of extremely metal-deficient stars (either lower-mass Population III or top-heavy Population II). 
This interpretation is physically viable; for instance our models of an evolved Population III cluster, where the most massive stars ($>40$\,\Msun) have already exploded, successfully reproduce the observed spectroscopic characteristics of LAP1-B (see Sect.~\ref{ssec:IMFs}).
A more complex history involving a second burst of Population III stars in re-accreted pristine gas is also theoretically possible \citesupp{wise2012, pallottini2014}. Regardless of the precise scenario (Population III or extreme Population II), the properties of LAP1-B require a mode of star formation dominated by exceptionally massive, metal-deficient stars, a process characteristic of the primordial Universe.

The existence of such a chemically primitive system at a relatively late cosmic epoch ($z=6.6$) is remarkable and requires explanation. Two primary scenarios can account for this. One possibility is that LAP1-B resides in a highly isolated, low-mass halo that only recently crossed the threshold for star formation. A more compelling explanation, however, involves the suppression of early star formation by a strong external Lyman-Werner (LW) radiation field. In this scenario, intense UV radiation dissociates molecular hydrogen, the primary coolant in pristine minihalos, thereby delaying the onset of star formation until the halo grows massive enough to cool via atomic lines. Recent cosmological simulations confirm that a high LW background can allow Population III star formation to persist down to $z\sim 6$ \citesupp{visbal2025} (see also ref \citesupp{zier2025} and references therein). The proximity of the LAP1-A complex may provide a natural source for the requisite intense LW radiation, making this a particularly attractive explanation for the delayed formation of LAP1-B.
\\

\Lya\ emission is clearly detected from LAP1-B, with a rest-frame equivalent width lower limit of $>250$\,\AA\ ($3\sigma$), interestingly with a sign of spatially extended.
The intrinsic equivalent width is likely to be even larger, given that LAP1-B lies within the epoch of reionization, where the surrounding circumgalactic and  intergalactic medium can partially attenuate escaping \Lya\ photons. Such a high EW(\Lya) is consistent with a metal-poor stellar population, although the current limit is not sufficient to uniquely identify a population of zero-metallicity stars \cite{inoue2011_metal_poor}.
The \Lya\ emission peak is redshifted relative to the systemic redshift determined from \Ha\ (Fig.~\ref{fig:spec_snapshot}b), with an observed velocity offset of $\Delta v_{{\rm Ly}\alpha} = 150 \pm 40$\,\kms. Such a redshifted \Lya\ is a common feature among high-redshift star-forming galaxies, and is typically attributed to radiative transfer effects arising from the resonant nature of \Lya\ photons, gas kinematics, and the column density of neutral hydrogen \citesupp{ouchi2020}. The measured offset in LAP1-B is small, falling within the typical range observed for the most intense \Lya-emitting galaxies at $z=2$--$3$ (i.e. after the end of reionization).
The detection of \Lya\ at $z=6.6$ with this modest velocity offset suggests a relatively low column density of neutral hydrogen ($N_{\mathrm{HI}}\lesssim 10^{19}$\,cm$^{-2}$) \citesupp{hashimoto2015,ouchi2020} both within the galaxy and in its surrounding circumgalactic and intergalactic environments. Such conditions are consistent with an efficient production of ionizing photons by a young, metal-poor stellar population, some of which may escape from the galaxy and contribute to cosmic reionization \citesupp{izotov2016_nature, vanzella2019_ion2, nakajima2020}.

\subsection*{Estimations of Stellar Mass, Gas Mass, and Dynamical Mass}
\label{ssec:masses_full}

The stellar mass of LAP1-B is constrained through the non-detection of continuum emission in deep JWST/NIRCam imaging of the MACS J0416 field. Broadband data covering this area are publicly available from the GTO programs PEARLS \citesupp{windhorst2023} and CANUCS \citesupp{willott2022}. 
We reprocessed the raw NIRCam data using the procedures described in ref \citesupp{harikane2023_first} and constructed composite images for each filter from F090W to F444W. LAP1-B remains undetected in all individual broadband images. To obtain the most stringent photometric limits, we created two stacked composite images: a ``blue'' composite (F115W, F150W, F200W), probing the rest-frame ultraviolet range of $1300$--$2900$\,\AA, and a ``red'' composite (F356W, F444W), covering the rest-frame optical continuum from $4100$ to $6500$\,\AA. No significant flux is detected in either composite. 
Using circular apertures with diameters of 0\farcs 10 and 0\farcs 26 (approximately twice the PSF FWHM), we derive $3\sigma$ upper limits of $31.51$ and $30.69$\,mag for the blue and red composites, respectively.
After correcting for lensing magnification of $\mu=98$ \cite{vanzella2023_metalpoor}$^{,}$\citemeth{bergamini2023}, the blue composite non-detection corresponds to an absolute UV luminosity limit of \Muv\ $>-10.4$ ($3\sigma$), confirming the ultra-faint nature of LAP1-B \cite{vanzella2023_metalpoor}.

We estimate the stellar mass upper limit using two independent approaches. First, we focus on the red composite, which probes the rest-frame optical continuum and is thus more sensitive to the accumulated stellar mass. However, this bandpass can be significantly contaminated by strong nebular emission lines such as \Ha, \Hb, and \OIII$\lambda5007$. We quantify the contribution of these lines by integrating their observed fluxes through the filter response curves of F356W and F444W, without applying a slit-loss correction. After subtracting this contamination, we estimate a corrected $3\sigma$ continuum limit of $31.12$\,mag. Adopting a stellar mass-to-light ratio appropriate for a young, metal-free Population~III population with a mass range of $1$--$100$\,\Msun\ \cite{schaerer2002}, this corresponds to an upper limit of $1.8 \times 10^4$\,\Msun\ for the stellar mass, after magnification correction. 
This optical limit is robust against changes in the assumed stellar population; adopting an extremely metal-poor Population II model ($Z=10^{-7}$, $1$--$100$\,\Msun\, 10 Myr) yields a consistent upper limit of $1.8 \times 10^4$\,\Msun.
Second, we use the blue composite, which probes the rest-frame UV ($\sim$2000\,\AA) and is more directly tied to ongoing star formation. While this wavelength range is less sensitive to total stellar mass, it provides more stringent constraints on the current starburst. Assuming the same Population~III model, the $3\sigma$ non-detection yields a mass limit of $<2,700$\,\Msun, also corrected for magnification. If we instead assume the aforementioned Population II model ($Z=10^{-7}$, 10 Myr), the UV-based mass limit increases only slightly to $<3,300$\,\Msun. To remain conservative and account for the uncertainty in the underlying stellar population, we adopt the latter value of \Mstar\ $<3,300$\,\Msun\ ($3\sigma$) as our fiducial upper limit, following the methodology of ref \cite{vanzella2023_metalpoor}.
Even taking our more conservative optically-derived limit ($<1.8 \times 10^4$\,\Msun) as a robust bound, the system remains within the same order of magnitude regardless of the chosen stellar population model. This confirms that LAP1-B is an extremely low-mass system, placing it firmly in the regime of the high-redshift progenitors of today's ultra-faint dwarf galaxies.
\\

To estimate the gas mass, we apply two independent approaches. 
First, we use the Kennicutt-Schmidt relation, which links the star formation rate (SFR) and gas mass surface density. This relation has been established for nearby star-forming galaxies \citesupp{kennicutt1998_KSLaw} and its validity has been tested at moderately high redshift \citesupp{bouche2007}. We estimate the SFR from the \Ha\ luminosity \citesupp{kennicutt1998}, corrected for lensing and scaled to a Chabrier IMF, consistent with the calibration used in ref \citesupp{bouche2007}. This method yields a star formation rate of SFR $= (5.3 \pm 0.6) \times 10^{-3}$\,\Msunyr.
To estimate the intrinsic size of the star-forming region, we use the spatial extent of the Hydrogen Balmer line map \cite{vanzella2023_metalpoor}, in combination with the lens model \citemeth{bergamini2023}, which constrains the semi-major axis of LAP1-B to be less than $20$\,pc. We adopt this as a conservative upper-limit, and consider alternative cases of $10$\,pc and $5$\,pc, comparable to the smallest sizes observed in present-day UFDs \cite{simon2019}, to explore the impact on the derived gas mass.
Assuming the gas is confined to the region traced by Balmer emission, the Kennicutt-Schmidt relation implies a scaling of $M_{\rm gas} \propto r^{0.83}$ using the form \citesupp{bouche2007}. For the $20$\,pc upper limit, we obtain $M_{\rm gas, 20\,pc} = 4.1^{+0.9}_{-0.3} \times 10^5$\,\Msun. For smaller sizes, the inferred masses are $M_{\rm gas, 10\,pc} = 2.3^{+0.5}_{-0.2} \times 10^5$\,\Msun\ and $M_{\rm gas, 5\,pc} = 1.3^{+0.3}_{-0.1} \times 10^5$\,\Msun. Taken together, these estimates constrain the total gas mass to lie within $\simeq (1$--$5) \times 10^5$\,\Msun, exceeding the stellar mass limit by two orders of magnitude. This strongly supports the interpretation that LAP1-B is a gas-dominated system with minimal stellar mass content. 
Adopting the conservative limit on the stellar mass does not change this conclusion.

As a second, complementary approach, we estimate the gas mass based on the observed oxygen abundance. Assuming metal production by a Population III, metal-free stellar population following a Salpeter IMF \cite{schaerer2002}, where stars with $M>10$\,\Msun\ contribute a typical oxygen yield of $\sim 0.1$\,\Msun\ per 1\,\Msun\ of stars formed/exploded (by referring to the metal production table \citesupp{nomoto2013}), we calculate the total oxygen mass produced. These massive stars are expected to comprise $\sim 10$\,\%\ of the stellar population in mass, implying a total oxygen mass of $\lesssim 30$\,\Msun\ ($3\sigma$), given the stellar mass limit. Based on the measured oxygen abundance of \Oabundance\ $=6.3$, which corresponds to a gas-phase oxygen mass fraction of $\sim 6\times 10^{-5}$, this implies a gas mass of  $\lesssim 5\times 10^5$\,\Msun. 
This estimate aligns well with the value derived independently from the Kennicutt-Schmidt relation. This consistency holds even when adopting our conservative upper limit on the stellar mass, which is a factor of $\sim 6$ higher.
\\

The dynamical mass of LAP1-B is evaluated from emission-line kinematics. The line profile of \Ha, which is identified at the highest signal-to-noise ratio in our spectrum, exceeds the instrumental broadening expected from the NIRSpec line spread function at this wavelength. A broader Gaussian profile provides a significantly better fit to the \Ha\ spectral profile than the instrumental broadening, yielding an intrinsic velocity dispersion of $58.3 \pm 17.8$\,\kms. While its gas velocity dispersion is higher than the stellar dispersions of UFDs \citesupp{mcconnachie2012}, this is expected, as the turbulent, star-forming gas in a high-redshift galaxy would be kinematically hotter than the relaxed stellar populations of a quenched relic.

We estimate the dynamical mass using the virial relation, $M_\mathrm{dyn} = C \sigma^2 r / G$, where $\sigma$ is the velocity dispersion and $C$ is a structural constant. Following previous work \citesupp{erb2006_mass}, we adopt $C = 3.4$, corresponding to the choice of $r$ as the effective radius of the \Ha-emitting region.
Using the size constraints discussed above, we derive dynamical masses of $M_{\rm dyn, 20\,pc} = 5.4^{+3.8}_{-2.8} \times 10^7$\,\Msun\ (upper limit), $M_{\rm dyn, 10\,pc} = 2.7^{+1.9}_{-1.4} \times 10^7$\,\Msun, and $M_{\rm dyn, 5\,pc} = 1.3^{+0.9}_{-0.7} \times 10^7$\,\Msun.
We note that this calculation assumes a constant velocity dispersion across the range of radii tested, a necessary simplification as our data do not spatially resolve the kinematic profile.
Assuming the baryonic mass is dominated by the gas component, the baryon fraction can be approximated as $M_{\rm gas}/M_{\rm dyn}$, which scales weakly with $r^{-0.17}$. We thus obtain baryon fractions of $\simeq 1$\%\ or less: specifically, $0.76$\%\ for $r = 20$\,pc, $0.85$\%\ for $r = 10$\,pc, and $1.0$\%\ for $r = 5$\,pc. These values indicate that the visible baryonic components are embedded within a dominant dark matter halo.

We can also quantify this dark matter dominance using the mass-to-light ($M_{\rm dyn}/L$) ratio. Our intrinsic $V$-band luminosity is limited to $M_V >-10.75$ ($3\sigma$), a value derived from the red composite image after correcting for both gravitational lensing and strong nebular line contamination. Combining this luminosity limit with our dynamical mass estimate yields a lower limit on the mass-to-light ratio of $M_{\rm dyn}/L_V \gtrsim 10$\,\MsunLsun. This value is fully consistent with the present-day UFD population, which exhibits $M_{\rm dyn}/L_V$ ratios spanning from $\sim 10$ to over $1000$ \cite{simon2019}$^{,}$\citesupp{mcconnachie2012, pace2025} across its faint luminosity range, placing LAP1-B squarely in the regime of dark matter-dominated systems.
We acknowledge that our dynamical mass estimate is subject to systematic uncertainties from assumptions about the system's structure. To quantify this, we explore how our estimate changes with different plausible assumptions. 
As a quantitative test, we apply the alternative prescription \citesupp{ubler2023_GA-NIFS}. In this framework, the structural coefficient, $C$, is calculated as the sum of two terms: $K(n)$, which is a function of the S\'ersic index ($n$) \citesupp{cappellari2006}, and $K(q)$, which is a function of the axis ratio ($q$)\citesupp{vanderWel2022}. By exploring a plausible range for these parameters (S\'ersic index $n=1-4$ and axis ratio $q=0-1$), we find that the coefficient $C$ can range from $4.6$ to $12.6$. This corresponds to a dynamical mass that is $1.3$ to $3.7$ times higher than our fiducial estimate.
Crucially, even when adopting the lowest dynamical mass in this range, it remains significantly larger than the baryonic mass. This test therefore reinforces our primary conclusion that LAP1-B is an overwhelmingly dark matter-dominated system.

\subsection*{Abundance Calibrations and Comparison with Present-day UFDs}
\label{ssec:UFDs_full}

To place LAP1-B in context, we compare its properties to those of ultra-faint dwarf galaxies (UFDs) in the present-day Universe. Our comparison sample is primarily drawn from Milky Way satellites, which represent the largest and most homogeneously studied population, but also includes UFDs from the environment of the Large Magellanic Cloud (LMC). Although the census in this external environment is less complete, their fundamental properties (e.g., luminosities and physical sizes) are broadly consistent with those of Milky Way UFDs \cite{simon2019}, justifying the use of a combined reference sample.

A primary challenge in this comparison is the conversion between gas-phase oxygen abundance and the stellar iron abundances ([Fe/H]) typically measured in local relics. Because direct oxygen measurements in UFDs are extremely limited \citesupp{ji2018,ji2020}, we adopt two distinct methodologies tailored to our primary diagnostic comparisons.
\\

\noindent
\textbf{Baseline [O/Fe] for the Mass-Metallicity Relation:} 
To compare LAP1-B with the general UFD population on the mass-metallicity relation (Fig.~\ref{fig:MZR}), we derive a representative [O/Fe] ratio based on Milky Way halo stars. This is motivated by the observation that $\alpha$-elements other than oxygen show similar abundance trends between UFD stars and the Milky Way halo at comparable metallicities \cite{simon2019} \citesupp{frebel2010}. This suggests that oxygen, another $\alpha$-element, likely follows a similar pattern. We utilize metal-poor halo stars from the SAGA database \citesupp{suda2008} as a proxy to determine a representative [O/Fe] value. From this database, we select stars with [Fe/H] $<-1$, excluding red giants (to avoid internal mixing effects \citesupp{gratton2000}) and CEMP stars. CEMP stars are excluded from this specific baseline as they often reflect highly specific or localized enrichment histories, such as binary mass transfer or pollution by specific supernovae, that place them well outside the standard abundance patterns \citesupp{BC2005}. From this curated ``normal'' metal-poor sample, we find that [O/Fe] exhibits a well-defined plateau at approximately $+0.5$ across the metallicity range relevant to UFDs.

To validate this assumption, we compare the directly measured [C/O] ratios of these halo stars with those inferred using our adopted conversion ([C/O] = [C/Fe] $-0.5$). As shown in Extended Data Fig.~\ref{fig:co_oh_MWhalostars}, the resulting distributions are in excellent agreement. We therefore adopt [O/Fe] $=+0.5$ as our baseline for the general UFD population. 
Beyond providing a robust calibration, Extended Data Fig.~\ref{fig:co_oh_MWhalostars} reveals a distinct chemical trend: as oxygen decreases toward the metal-poor regime, the Galactic halo stars exhibit an upward turn in C/O, a feature also highlighted in the broader context of Fig.~\ref{fig:co_oh} (see below).
While a few individual UFD stars show higher [O/Fe] ratios approaching $+1.0$ \citesupp{ji2018, ji2020}, these likely represent observational selection biases toward O-rich outliers. We maintain that our statistically robust [O/Fe] $=+0.5$ baseline remains the most reasonable choice for the average UFD population.

In Fig.~\ref{fig:MZR}, we plot the relationship defined by local dwarf galaxies \citesupp{pace2025}, which utilizes the local volume database \citesupp{pace2024}. Following standard classifications, we distinguish between UFDs, defined by stellar masses \Mstar\ $\leq 10^5$\,\Msun, and classical dwarf spheroidal galaxies (dSphs) with \Mstar\ $>10^5$\,\Msun\ \citesupp{kirby2008}. For this sample, stellar masses are derived from integrated $V$-band magnitudes assuming a standard mass-to-light ratio of 2 \citesupp{mcconnachie2012}, while their metallicities are based on the metallicity distribution of member stars with the assumption of [O/Fe] $=+0.5$. The dotted line represents a simple linear fit to the combined sample of UFDs and dSphs, including satellites of both the Milky Way and the LMC. We adopt this linear representation, following recent literature \citesupp{pace2025}, as a simple visual guide to the overall trend in the current data, while noting the potential physical expectation of a flattening at the lowest masses \citesupp{sanati2023, go2025, wheeler2025}.
LAP1-B sits at the low-mass extreme of this relation, specifically occupying the regime of the most metal-poor UFDs associated with the LMC \citesupp{pace2025}.
\\

\noindent
\textbf{Metallicity-dependent [O/Fe] for CEMP stars:} 
For the carbon-to-oxygen comparison (Fig.~\ref{fig:co_oh}), we shift our focus from the ``average'' population to individual CEMP stars identified in UFDs. These stars are widely interpreted as the direct fossil record of the earliest star-formation events.
We select stars with [C/Fe] $\gtrsim 1.0$ from the compilation of UFD stars presented in \citesupp{pace2025}. Note that the UFD's stellar abundances are all corrected for evolutionary status \citesupp{placco2014}. For these objects, we avoid a constant [O/Fe] and instead utilize the empirical distribution of CEMP-no stars, i.e. CEMP stars with no s-process enhancement, in the Galactic halo, which reflect natal gas composition rather than binary pollution, to derive a metallicity-dependent [O/Fe] conversion. Using the CEMP-no stars in the halo exploited from the SAGA database, we derive a best-fit relation: [O/Fe] $=-1.7\times$ [Fe/H] $-4.5$, with a $\pm 0.3$\,dex uncertainty (68th percentile). This relation is applied to UFD CEMP stars within the relevant metallicity range ([Fe/H] $=-4$ to $-2.5$). In Fig.~\ref{fig:co_oh}, the [O/Fe] uncertainty is represented by diagonal error bars, accounting for the coupled nature of the [C/O] and [O/H] conversion. This refined analysis shows that a subset of CEMP stars in UFDs possesses C/O and O/H ratios consistent with LAP1-B.
\\

While our measurement provides a snapshot of the gas-phase abundance and the UFD data represent a quenched, mass-averaged relic, the fundamental similarity is striking. The gas we observe is the raw material from which the bulk of a UFD's stars would eventually form. The expected locus of LAP1-B as a final stellar population in the mass-metallicity diagram is somewhat complex, owing both to potential further enrichment beyond its observed stage and to an offset introduced by averaging over earlier, more metal-poor stellar populations. Crucially, however, we find that LAP1-B's properties are already consistent with those of local UFDs without invoking any such evolution. 
LAP1-B is therefore a compelling candidate for a direct UFD progenitor, observed during its brief, primary star-formation episode.

\subsection*{Robustness tests: The effect of IMF and stellar metallicity}
\label{ssec:IMFs}

Our fiducial model for the Population III interpretation is based on an IMF with a Salpeter slope over a $1$--$100$\,\Msun\ mass range, which we consider a conservative baseline. However, as both purely theoretical \citesupp{hirano2015} and data-constrained \citesupp{koutsouridou2024, ferrara2026} models suggest the presence of significantly more massive Population III stars, potentially favouring a more top-heavy IMF, we now explore how our results are affected by alternative stellar population assumptions.
\\

To investigate a top-heavy scenario, we adopt the Population III model \cite{schaerer2002}, which considers stars in the mass range of $50-500$\,\Msun. The resulting line ratios are shown in Extended Data Fig.~\ref{fig:models_popIII_topheavyIMF}. While the predicted \CIV/\OIII\ ratio is slightly higher than in our baseline model, the shift is minor, and our main conclusion persists: the observed line ratios are well reproduced by photoionization from a zero-metallicity population with a C/O ratio of $1$--$2\times$ solar (Equation~\ref{eq:co}). The predicted \OIII/\Hb\ ratios are also nearly indistinguishable from our fiducial model (Extended Data Fig.~\ref{fig:Z_R3}), resulting in an oxygen abundance that is only $0.05$\,dex higher, well within the measurement uncertainty.
Regarding the stellar mass, this top-heavy IMF has a lower mass-to-light ratio, which leads to new upper limits of $600$\,\Msun (from the rest-frame UV) and $5,300$\,\Msun (from the optical), after correcting for magnification. While stringent, these are consistent with our original limit of $3,300$\,\Msun. Therefore, our conclusion that LAP1-B is an extremely low-mass, gas-dominated system remains robust.
\\

Alternatively, it is possible that the most massive Population III stars have already died, producing a small amount of metals that slightly enrich the ISM. In this scenario, the currently shining population would lack the highest-mass stars. To explore this, we compute a new model with a truncated IMF spanning $1$--$40$\,\Msun\ using the code \citemeth{gessey-jones2022}.
The predicted \CIV/\OIII\ and \OIII/\Hb\ line ratios from this truncated model are shown in Extended Data Fig.~\ref{fig:models_popIII_bottomheavyIMF} which are nearly identical to our baseline, reinforcing our conclusions about the ionization parameter and metallicity. As expected, removing the most luminous stars increases the mass-to-light ratio. This leads to revised stellar mass upper limits of $<2,900$\,\Msun\ (UV) and $<1.9\times 10^4$\,\Msun\ (optical). This new UV-based limit is comparable to our fiducial value, confirming that our low-mass estimate is not significantly affected by the absence of the most massive stars.
\\

Finally, we explore alternative Population II models to determine if an extremely metal-poor, top-heavy IMF could explain LAP1-B's hard ionizing spectrum. For this, we adopt the $Z=10^{-7}$ ($\simeq 0.0007$\% solar) Population II models \cite{schaerer2003} and explore two IMFs: a standard Salpeter ($1$--$100\,M_{\odot}$) and a massive-only ($50$--$500\,M_{\odot}$) case. The results are shown in Fig.~\ref{fig:c4o3_o3hb} (bottom) and Extended Data Fig.~\ref{fig:diagnostics_popIII}. The \CIV/\OIII\ diagnostic is highly sensitive to depletion assumptions; while all Population II models (including the fiducial ones) fail to reproduce the observed ratio with standard depletions, their non-depleted versions can match the data (Fig.~\ref{fig:c4o3_o3hb}, bottom right). However, the \xiion\ measurement provides a more robust constraint on the ionizing source. As shown in Extended Data Fig.~\ref{fig:diagnostics_popIII}, the large \xiion\ of LAP1-B can only be explained by either the Population III models or the extreme $Z=10^{-7}$ model with the $50$--$500\,M_{\odot}$ IMF. These comparisons thus rule out all standard stellar populations and provide a compelling alternative interpretation: LAP1-B is powered by an extremely metal-deficient, massive stellar population.


\bibliographystylesupp{sn-mathphys-num}
\bibliographysupp{Refs_sn-article}{}


\end{document}